\documentclass[useAMS,usenatbib,a4paper]{mn2e}
\usepackage{graphicx}
\usepackage[fleqn]{amsmath}
\usepackage{amssymb,times}

\defcitealias{pm05}{PM05}
\defcitealias{rj95}{RJ95}
\title[A switch to reduce resistivity in SPMHD]{A switch to reduce resistivity in smoothed particle magnetohydrodynamics}
\author[Tricco \& Price]{Terrence S. Tricco$^1$\thanks{email: terrence.tricco@monash.edu} and Daniel J. Price$^1$\\$^1$Monash Centre for Astrophysics and School of Mathematical Sciences, Monash University, Clayton, Victoria, 3800, Australia}

\begin{document}

\maketitle

\begin{abstract}
Artificial resistivity is included in Smoothed Particle Magnetohydrodynamics simulations to capture shocks and discontinuities in the magnetic field.  Here we present a new method for adapting the strength of the applied resistivity so that shocks are captured but the dissipation of the magnetic field away from shocks is minimised.  Our scheme utilises the gradient of the magnetic field as a shock indicator, setting $\alpha_\text{B} = h \vert \nabla {\bf B} \vert / \vert {\bf B} \vert$, such that resistivity is switched on only where strong discontinuities are present.  The advantage to this approach is that the resistivity parameter does not depend on the absolute field strength.  The new switch is benchmarked on a series of shock tube tests demonstrating its ability to capture shocks correctly.  It is compared against a previous switch proposed by \citet{pm05}, showing that it leads to lower dissipation of the field, and in particular, that it succeeds at capturing shocks in the regime where the Alfv\'en speed is much less than the sound speed (i.e., when the magnetic field is very weak). It is also simpler. We also demonstrate that our recent constrained divergence cleaning algorithm has no difficulty with shock tube tests, in contrast to other implementations.
\end{abstract}


\section{Introduction}
Magnetised shocks and discontinuities pervade the interstellar medium \citep{es04,gaensleretal11}. Capturing these properly in numerical simulations is critical to accurately predicting the formation of stars from turbulent, magnetised, molecular clouds \citep{fk12}. On the other hand, estimates of the microscopic viscosity and resistivity in the interstellar medium suggest very high values of the kinematic and magnetic Reynolds numbers, respectively, typically orders of magnitude higher than can be achieved in numerical codes \citep[c.f.][]{es04}. Thus, it is important to minimise numerical dissipation in simulation codes.

Magnetic fields in astrophysics can be described with the equations of magnetohydrodynamics (MHD), given here in the ideal (non-dissipative) limit as,
\begin{align}
\frac{{\rm d}\rho}{{\rm d}t} & = -\rho \nabla \cdot {\bf v}, \label{eq:cty} \\
\frac{{\rm d}{\bf v}}{{\rm d}t} & =  -\frac{1}{\rho}\nabla\left(P + \frac{B^2}{2 \mu_0} - \frac{{\bf B B}}{\mu_0}\right), \label{eq:mom} \\
\frac{{\rm d} {\bf B}}{{\rm d}t} & =  \left( {\bf B} \cdot \nabla \right) {\bf v} - {\bf B} \left( \nabla \cdot {\bf v} \right). \label{eq:ind} 
\end{align}
where ${\rm d}/{\rm d}t \equiv \partial  / \partial t + {\bf v} \cdot \nabla$, $\rho$ is the density, ${\bf v}$ is the velocity, $P$ is the pressure and ${\bf B}$ is the magnetic field.

The basic procedure for solving the MHD equations in the context of the Smoothed Particle Hydrodynamics (SPH) method was developed by \citet{phillipsmonaghan85} and \citet{pm04a,pm04b,pm05}, as summarised in the recent review by \citet{price12}. The main issues are: i) removing the tensile instability, achieved by subtracting the $-{\bf B}(\nabla\cdot{\bf B})/\rho$ source term from Eq.~\ref{eq:mom} as proposed by \citet{borve01}; ii) to formulate dissipative terms for capturing shocks and other discontinuities \citep{pm04a,pm05,price08}; and iii) maintaining the solenoidal constraint on the magnetic field. We have recently addressed the last issue by formulating a constrained version of \citet{dedneretal02}'s hyperbolic/parabolic divergence cleaning algorithm \citep{tricco&price12}, avoiding problems associated with earlier approaches involving the Euler potentials \citep{pricebate07,rosswogprice07} or the vector potential \citep{price10}. Here we address issue ii) in more detail.
 
The usual approach to shock-capturing in SPH \citep[see reviews by][]{monaghan05,price12} is to treat discontinuities in fluid variables by adding dissipation terms which smooth the variable across sharp jumps in order to resolve the discontinuity. Artificial viscosity for treatment of hydrodynamic shocks was developed by \citet{mg83}.  In this paper, we use the form of artificial viscosity by \citet{monaghan97}, developed by analogy with Riemann solvers, giving an additional term in Eq.~\ref{eq:mom} of the form
\begin{equation}
 \left(\frac{{\rm d}{\bf v}_a}{{\rm d}t}\right)_{\rm visc} = \sum_b m_b \frac{\alpha v_\text{sig}}{\overline{\rho}_{ab}} ({\bf v}_a - {\bf v}_b) \cdot \hat{{\bf r}}_{ab} \nabla_a W_{ab}.
\label{eq:visc}
\end{equation}
This is an SPH representation, using the gradient of the smoothing kernel $W$, of a viscosity term with dissipation parameter $\nu \propto \alpha v_{\rm sig} h$. The parameter $\alpha$ is dimensionless and of order unity. The characteristic velocity of the shock is represented using the signal velocity, $v_\text{sig} = 0.5 (c_a + c_b - \beta {\bf v}_{ab} \cdot \hat{{\bf r}}_{ab})$ with $\beta\sim2$, which is the maximum speed of information propagation between each pair of particles. The dissipation term is thus first order with respect to the resolution length, $h$ (hence `artificial' rather than physical viscosity, because the diffusion parameter is proportional to resolution). \citet{monaghan97} also proposed an artificial thermal conductivity term that turns out to be important in simulating contact discontinuities, where incorrect treatment can affect the development of Kelvin-Helmholtz instabilities \citep{price08, wadsley08}. For a full discussion on discontinuities in SPH, see \citet{price08, price12}.

To reduce unwanted dissipation away from discontinuities, \citet{mm97} allowed $\alpha$ to be spatially variable, using a switch so that $\alpha\to 1$ only in the presence of shocks.  In their work, $\alpha$ for a given particle, $a$, is integrated according to
\begin{equation}
 \frac{{\rm d} \alpha_a}{{\rm d}t} = \max(- \nabla \cdot {\bf v}_a, 0) - \frac{\alpha_a - \alpha_\text{min}}{\tau} ,
\end{equation}
where $\tau = h v_\text{sig} / C$, $h$ is the smoothing length, and $C\sim0.1$ corresponds to a decay scale of approximately 5 smoothing lengths to the minimum $\alpha_\text{min}=0.1$. A limiter to suppress viscosity in the presence of shear flows was also introduced by \citet{balsara95}.  Recently, several authors have proposed improved $\alpha$ viscosity switches to improve shock detection while reducing dissipation away from shocks.  \citet{cd10} suggest monitoring ${\rm d} / {\rm d} t ( \nabla \cdot {\bf v})$ as the shock indicator which they find activates $\alpha$ earlier when a shock is approaching, and leads to less overall dissipation.  \citet{rh12} proposed a similar approach using $\nabla (\nabla \cdot {\bf v})$.

In SPMHD, an artificial resistivity for the magnetic field is included to capture magnetic shocks and discontinuities (i.e., current sheets). The standard implementation of \citet{pm05} (hereafter \citetalias{pm05}) adds a term to the induction equation of the form
\begin{equation}
 \left( \frac{{\rm d}{\bf B}_a}{{\rm d}t} \right)_\text{diss} = \rho_a \sum_b m_b \frac{\overline{\alpha}_{B,ab} v^{B}_\text{sig}}{\overline{\rho}_{ab}^2} ({\bf B}_a - {\bf B}_b) \hat{r}_{ab} \cdot \nabla_a W_{ab} ,
\label{eq:resistivity}
\end{equation}
where $\alpha_\text{B}$ is similarly a dimensionless quantity of order unity and $v^B_\text{sig}$ is a signal velocity. As for artificial viscosity, this is simply a standard representation of a diffusion term in SPH \citep[see e.g.][]{monaghan05,price12} but where the diffusion parameter $\eta \propto \alpha_\text{B} v^{B}_{\rm sig} h$. However, the choice of signal velocity in this case is less clear. Ideal MHD has three wave solutions but without reconstructing the full Riemann state it is not possible to determine the type of shock.  Thus, this is typically chosen to be the speed of the fast MHD wave. Since this is rather dissipative, \citet{price12} instead suggested using the averaged Alfv\'en velocity as the choice of signal velocity.

Similar to viscosity, a switch may be employed for $\alpha_\text{B}$ to reduce dissipation away from shocks. By analogy with \citet{mm97}, \citetalias{pm05} suggested using
\begin{equation}
\label{eq:intalphab}
 \frac{{\rm d}\alpha_{\text{B},a}}{{\rm d}t} = \max\left( \frac{\vert \nabla \cdot {\bf B}_a \vert}{\sqrt{\mu_{0}\rho_a}}, \frac{\vert \nabla \times {\bf B}_a \vert}{\sqrt{\mu_{0}\rho_a}}\right) - \frac{\alpha_{\text{B},a}}{\tau} .
\end{equation}
This switch works satisfactorily for many problems, leading to sharper jump profiles and a decrease in the overall dissipation of the magnetic field.  However, \citet*{ptb12} noted in their star formation simulations that, even with this switch, excess dissipation could suppress the formation of protostellar jets.

Our need for a new resistivity switch is motivated by the failure of the \citetalias{pm05} switch in the limit where the Alfv\'en speed is much smaller than the sound speed, as will be shown in section \ref{sec:mhdturb}.  Since $\alpha_\text{B} \propto \vert \nabla \times {\bf B}\vert $ (assuming $\nabla \cdot {\bf B}$ is negligible), this means that $\alpha_\text{B}$ is related to the magnitude of the magnetic field.  Thus, for weak fields $\alpha_\text{B}$ may remain quite small even in the presence of strong shocks.

In this work, we present a new switch for $\alpha_\text{B}$ that captures shocks in the magnetic field in both weak and strong fields.  This addresses the deficiencies of the previous switch and results in less overall dissipation of magnetic energy. The paper is organised as follows: In Sec.~\ref{sec:formulation}, the new resistivity formulation and implementation is described. Sec.~\ref{sec:tests} contains a suite of tests designed to test the efficacy of the new switch and to compare results against the previous switch. Results are summarised in Section~\ref{sec:conclusion}.

\section{Formulation}
\label{sec:formulation}
Our approach is to utilise $\nabla {\bf B}$, the $3\times3$ gradient matrix of ${\bf B}$, as the shock indicator.  For each particle, $\alpha_\text{B}$ is directly set to the dimensionless quantity
\begin{equation}
\alpha_{\text{B},a} = \frac{h_a \vert \nabla {\bf B}_a \vert}{\vert {\bf B}_a \vert} ,
\label{eq:alpha_B}
\end{equation}
which is restricted to the range $\alpha_\text{B} \in [0,1]$. 

By using the norm of the gradient of the magnetic field normalised by the magnitude of the magnetic field, the dependence on magnetic field strength is removed and this gives a relative measure of the strength of the discontinuity.  This allows shocks and discontinuities to be robustly detected in both the weak and strong field regimes.  It naturally produces values of $\alpha_\text{B}$ in the desired range and of the appropriate size for the discontinuity encountered, with regions away from shocks having negligible $\alpha_\text{B}$ values. 
% It also has no temporal dependence, unlike the PM05 switch.

The numerical dissipation of the magnetic field should scale quadratically with resolution when using this switch.  Artificial resistivity without using a switch scales linearly with resolution, which is evident from Eq.~\ref{eq:resistivity}.  The new switch adds an additional linear scaling with $h$ hence, in principle, quadratic scaling should be obtained.

The switch produces the same $\alpha_\text{B}$ values for multiplicative increases in magnetic field strength, important for dynamo-type problems where the magnetic field grows in strength. This represents a significant advantage over the \citetalias{pm05} switch.  Additive increases to the magnetic field, however, will yield different values of $\alpha_\text{B}$, and using this switch in relativistic contexts would require further consideration.

An obvious issue is what happens when $\vert {\bf B} \vert \to 0$.  This situation occurs in current sheets or null points where the magnetic field undergoes a reversal in direction.  In these cases, $\alpha_\text{B} \to 1$, which is the correct behaviour for current sheets since they represent a discontinuity in the magnetic field, but is not so for null points.  Dividing by zero can be avoided by adding a small parameter $\epsilon$ to $\vert {\bf B} \vert$.

\subsection{Implementation}

Each component of the gradient matrix is estimated using a standard SPH first derivative operator \citep[e.g.][]{price12},
\begin{equation}
\nabla{\bf B}_{a} \equiv \frac{\partial B^i_{a}}{\partial x^j_{a}} \approx -\frac{1}{\Omega_a \rho_a} \sum_b m_b ({B}^{i}_a - {B}^{i}_b) \nabla_a^j W_{ab}(h_{a}) ,
\label{eq:sphgradb}
\end{equation}
where $\Omega_a$ accounts for variable smoothing length terms.  This operator yields an estimate which is exact for constant functions.  We also investigated using an operator that is exact for linear functions, which may be obtained by performing a Taylor series expansion about ${\bf r}_a$ and solving a matrix inversion of the second error term \citep[see][]{price12}.  However, no difference was found for any of the tests shown in this paper, suggesting that this is unnecessary.

The norm of $\nabla {\bf B}$ is calculated using the 2-norm,
\begin{equation}
\vert \nabla {\bf B} \vert \equiv \sqrt{ \sum_i \sum_j \left\vert  \frac{\partial B^i_{a}}{\partial x^j_{a}} \right\vert^2 } .
\end{equation}
Several choices for computing this norm were investigated, such as the 1-norm, but no significant differences were found.

We investigated using the curl of the magnetic field as the shock indicator. While tests found it to be just as effective at detecting isolated shocks, we found that it did not measure discontinuities as well as the full gradient in complicated shock interactions.  The full gradient has further advantage in that the trace of the matrix produces the divergence of the field, meaning that dissipation will be applied if large divergence errors are present.

Finally, a \citet{cd10}-like approach was also investigated, whereby a time-dependent decay term for $\alpha_\text{B}$, similar to that in Eq.~\ref{eq:intalphab}, was added.  In this case, $\alpha_\text{B}$ was set using Eq.~\ref{eq:alpha_B} whenever this exceeded the current value, otherwise the existing value was retained and subsequently reduced on the next integration timestep using the decay term.  The aim was to let $\alpha_\text{B}$ smoothly decay after a shock had passed to improve representation of the post-shock field. We found that using Eq.~\ref{eq:alpha_B} alone already gives a smooth distribution in $\alpha_\text{B}$ about the centre of the shock, indicating that a decay term is not necessary for resistivity.

\subsection{Choice of signal velocity}
Similar to \citet{price12} we take the signal velocity to be an average of the wave speeds between the two particles
\begin{equation}
 v^{B}_\text{sig} = 0.5 (v_a + v_b),
\end{equation}
where $v$ is an appropriate MHD wave speed. The $-\beta {\bf v}_{ab} \cdot \hat{{\bf r}}_{ab}$ term used in the viscosity signal velocity, which corrects for the relative velocity of the particles and prevents particle interpenetration, is not included.  We find that for resistivity it is unnecessary and causes excessive dissipation. It may be noted that use of the averaged Alfv\'en speed for a signal velocity by \citet{price12} also excluded this term.

Unlike \citet{price12}, we find that the best choice is to use the fast MHD wave speed, as in the original \citet{pm04a} formulation, such that
\begin{equation}
 v_{a}^{2} = \frac{1}{2} \left(c^2_a + v_{A,a}^2\right) + \frac12 \sqrt{ (c_{a}^2 + v_{A,a}^2)^2 - 4 c_{a}^2 v_{A,a}^2 (\hat{{\bf B}_a} \cdot \hat{{\bf r}}_{ab}) },
\label{eq:vsigfastmhd}
\end{equation}
which is a composition of the sound speed, $c$, and the Alfv\'en speed, $v_A = B/ \sqrt{\mu_0 \rho}$.  If $c \gg v_A$, we find that \citet{price12}'s suggestion to use the Alfv\'en speed in the applied resistivity is insufficient to capture fast wave shocks (see Sec.~\ref{sec:mhdturb}).  When $v_A \gtrsim c$, the Alfv\'en speed and the fast wave speed will differ by less than a factor of 2.

\subsection{Switches using a second derivative}

In principle, a switch constructed using a higher derivative should provide a more reliable measure of the presence of a discontinuity in the magnetic field.  One suggestion by the referee of this paper, Walter Dehnen, is to use $\alpha_\text{B} = h \vert \nabla^2 {\bf B} \vert / \vert \nabla {\bf B} \vert$. Another option could be $\alpha_\text{B} = h^2 \vert \nabla^2 {\bf B} \vert / \vert {\bf B} \vert$, which would scale quadratically with resolution. 

The main difficulty in implementing higher derivative switches is calculating the second derivative in a way which is sufficiently free of noise from particle disorder.  We investigated calculating $\nabla^2 {\bf B}$ using the \citet{brookshaw85} form, that is, 
\begin{equation}
\nabla^{2} {\bf B}_{a} = \frac{2}{\Omega_a \rho_a} \sum_b m_b \left( {\bf B}_a - {\bf B}_b \right) \frac{F_{ab} (h_a)}{\vert r_{ab} \vert} ,
\end{equation}
where $\nabla W_{ab} \equiv ({\bf r}_{a} - {\bf r}_{b}) F_{ab}$, and also by taking two first derivatives as in Eq.~\ref{eq:sphgradb}, which, by taking two successive first derivatives, should lead to a more smooth estimate of the second derivative.  However, both of these simple estimates are significantly noisy when the particles are disordered, leading to high $\alpha_\text{B}$ and excessive dissipation.  The M6 quintic spline kernel was used in an attempt to reduce this noise, both by yielding a more regular particle distribution and a more accurate derivative estimate, but did not change the results.  

Therefore, a switch utilising the second derivative must use a higher order estimate in order to reduce noise from particle disorder, a conclusion similarly reached by \citet{cd10} and \citet{rh12}.  The most straightforward approach is to use two exact linear first derivatives which removes the $O(h)$ error term by taking a Taylor series expansion about ${\bf r}_a$ and performing a matrix inversion of the second error term.  Specifically, after first calculating $\nabla {\bf B}$ in such a manner, we compute
\begin{equation}
 \chi^{\sigma \gamma} \frac{\partial {\nabla {\bf B}}^{\alpha\beta}_a}{\partial {\bf x}^\sigma} = \sum_b m_b \left[ (\nabla {\bf B})_b^{\alpha \beta} - (\nabla {\bf B})_a^{\alpha \beta}\right] \nabla^{\gamma} W_{ab} 
\end{equation}
to obtain $\nabla^2 {\bf B}$, where $\chi^{\sigma\gamma} = \sum_b m_b ({\bf r}_b - {\bf r}_a)^\sigma \nabla^\gamma W_{ab}$ is the \mbox{$3\times 3$} matrix that must be inverted \citep[see][]{price12}.  This significantly improves the quality of the second derivative estimate, but requires two loops over the particles prior to the main loop where the resistivity term is calculated, meaning that it makes the overall SPMHD scheme $\sim1.5\times$ more expensive. This is a hefty price to pay for a switch that only marginally improves over Eq.~\ref{eq:alpha_B}. The second derivative evaluation proposed by \citet{rh12} is even more expensive, requiring a $10\times 10$ matrix inversion, and a minimum of 400 neighbours under the kernel. 

Our overall conclusion is to prefer the simple switch of Eq.~\ref{eq:alpha_B} for general use.  It performs robustly and effectively (see Sec.~\ref{sec:tests}), yet is simple to implement and cost-effective.

\section{Numerical Tests}
\label{sec:tests}

Our choice of tests are designed to study the ability of the switch to i) properly capture and model shock phenomena, and ii) suppress dissipation in areas away from shocks.  We have used three shocktube tests to study the former, using tests introduced by \citet{dw94} and \citet{briowu88} (corresponding to tests 1B, 2A, and 5A in \cite{rj95} (hereafter RJ95) whose naming convention we adopt).  These tests contain fast and slow shocks, fast and slow rarefactions, rotational discontinuities, and compound shock structures and are chosen to test the switch's ability to model all these separate shock types.  We then compare the new switch to the \citetalias{pm05} switch for three separate test problems: Propagation of a circularly polarised Alfv\'en wave, the Orszag-Tang vortex, and Mach 10 shocks in a fluid with an extremely weak field.  The last test is of particular interest because, as will be shown, the \citetalias{pm05} switch fails to recognise shocks in this weak field regime causing unphysical behaviour.

All our tests employ the constrained divergence cleaning algorithm of \citet{tricco&price12}.  The tests presented here serve to further validate this scheme.

\subsection{Shocktube 1B}
\label{sec:shock1b}

\begin{figure}
 \includegraphics[width=1.0\linewidth]{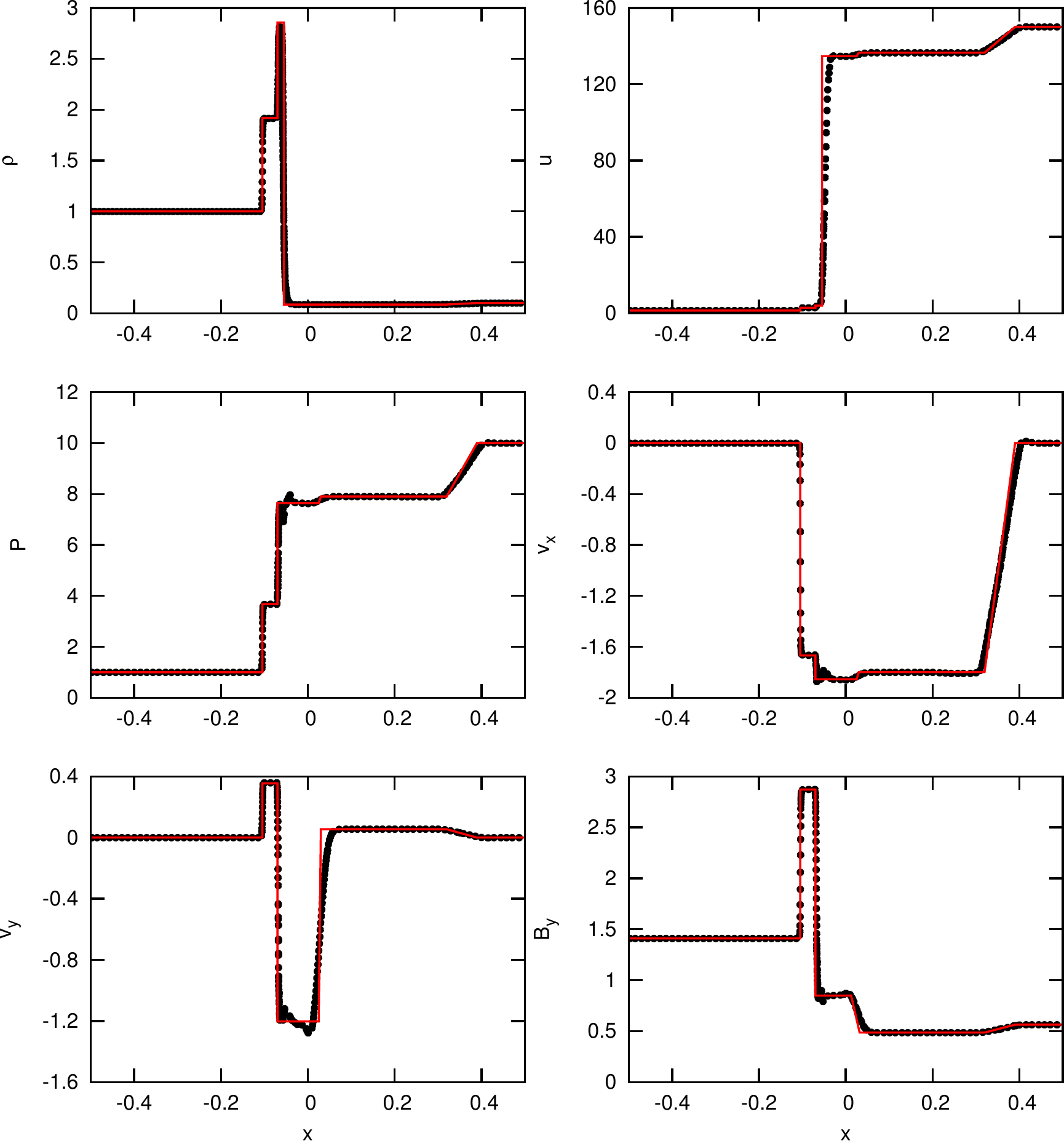}
\caption{Shocktube test 1B from \citetalias{rj95} performed in 2D with left state ($\rho$, $P$, $v_x$, $v_y$, $B_y$) $=$ (1, 1, 0, 0, $5/\sqrt{4 \pi}$) and right state ($\rho$, P, $v_x$, $v_y$, $B_y$) $=$ (0.1, 10, 0, 0, $2/\sqrt{4 \pi}$) with $B_x=3/\sqrt{4 \pi}$ at $t=0.03$.  Black circles are the particles and the red line is the solution from \citetalias{rj95}.}
\label{fig:shock1b}
\end{figure}

The first shocktube is a 2D test from \citet{dw94} which creates fast and slow shocks travelling in the -x direction, fast and slow rarefactions travelling in the +x direction, with a contact discontinuity in the centre.  The initial state for $x < 0$ (the `left state') is ($\rho$, $P$, $v_x$, $v_y$, $B_y$) $=$ (1, 1, 0, 0, $5/\sqrt{4 \pi}$), while for $x > 0$ (the `right state') is ($\rho$, P, $v_x$, $v_y$, $B_y$) $=$ (0.1, 10, 0, 0, $2/\sqrt{4 \pi}$) with $B_x=3/\sqrt{4 \pi}$ and $\gamma=5/3$.  

For this particular test, the initial density profile was used to calculate the initial thermal energy so that it forms a smooth transition between the two states.  This mitigates the presence of artificial pressure spikes in the initial conditions due to the high density contrast ($10$:$1$), seen also by \citet*{hubberetal13} in their studies of Kelvin-Helmholtz instabilities.

The shocktube has been simulated with 800$\times$26 particles for the left state and 260$\times$8 particles for the right state arranged on a triangular lattice.  Results at $t=0.03$ are shown in Fig.~\ref{fig:shock1b} and may be compared with the \citetalias{rj95} solution for the fast and slow shock and rarefactions (red line).   The L1 error in the $B_y$ profile is $8.911\times 10^{-3}$.  This compares to $9.547\times 10^{-3}$ if the shocktube is run using the \citetalias{pm05} switch.

For this shocktube, it is worth noting that no difficulties were found with our divergence cleaning algorithm.  Recently, \citet*{sdb12} published a different implementation and found that for this test it resulted in significant errors unless the cleaning was artificially limited.  This is due to the sharp $10:1$ density ratio that is unstable for their formulation, an issue that has been addressed and fixed in our cleaning algorithm without the need for artificial limiters \citep[see][]{tricco&price12}.

\subsection{Shocktube 2A}
\label{sec:shock2a}

\begin{figure}
 \includegraphics[width=1.0\linewidth]{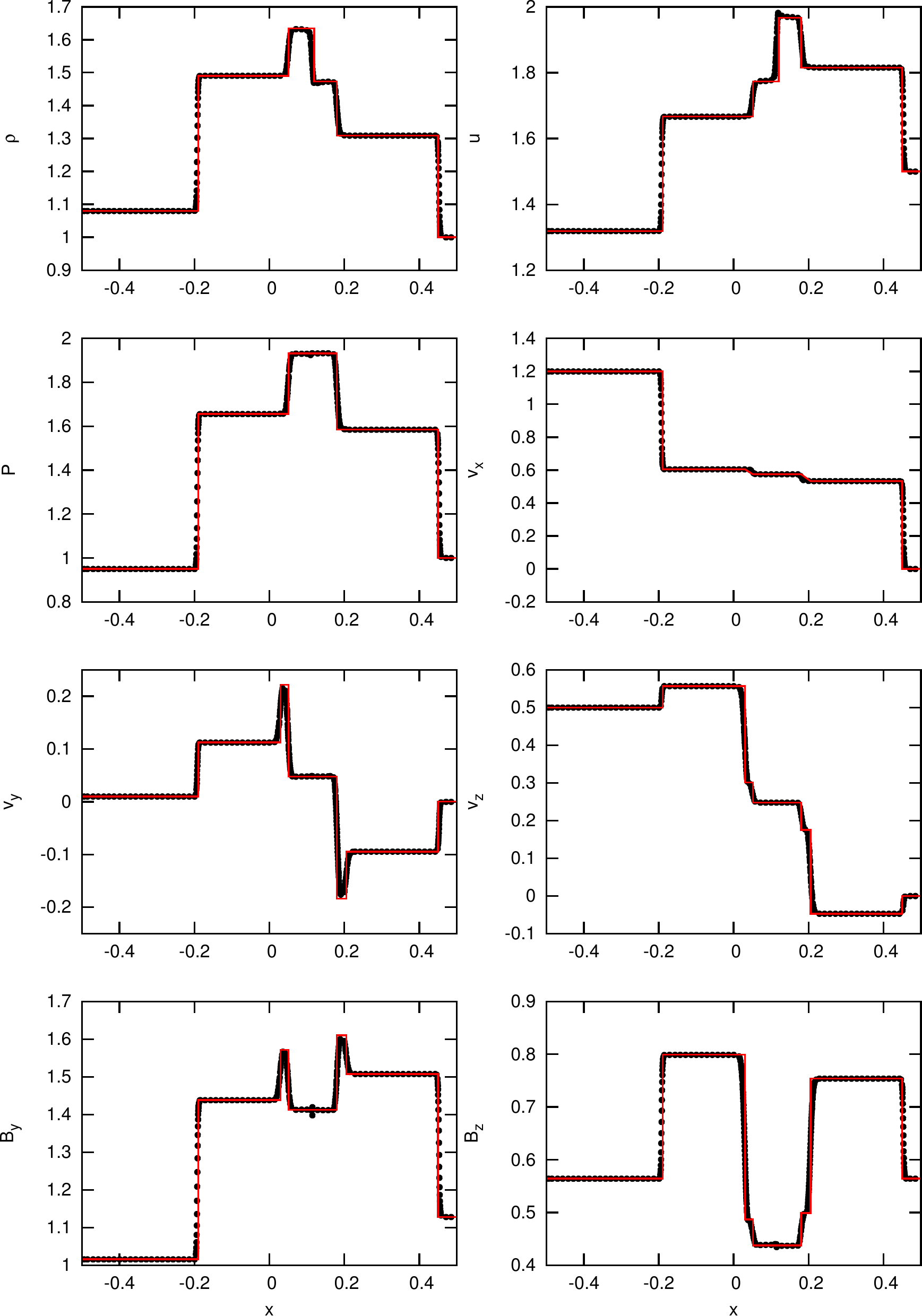}
\caption{Shocktube test 2A from \citetalias{rj95} performed in 3D with left state ($\rho$, $P$, $v_x$, $v_y$, $v_z$, $B_y$) $=$ (1.08, 0.95, 1.2, 0.01, 0.5, $3.6/\sqrt{4 \pi}$) and right state ($\rho$, P, $v_x$, $v_y$, $v_z$, $B_y$) $=$ (1, 1, 0, 0, 0, $4/\sqrt{4 \pi}$) with $B_x=B_z=2/\sqrt{4 \pi}$ at $t=0.2$. Black circles are the particles and the red line is the solution from \citetalias{rj95}.}
\label{fig:shock2a}
\end{figure}

This 3D problem originally introduced by \citet{dw94} has three dimensional velocity and magnetic fields generating two fast and slow shocks travelling in both directions, two rotational discontinuities, and a contact discontinuity in the centre.  It has left state ($\rho$, $P$, $v_x$, $v_y$, $v_z$, $B_y$) $=$ (1.08, 0.95, 1.2, 0.01, 0.5, $3.6/\sqrt{4 \pi}$) and right state ($\rho$, P, $v_x$, $v_y$, $v_z$, $B_y$) $=$ (1, 1, 0, 0, 0, $4/\sqrt{4 \pi}$) with $B_x=B_z=2/\sqrt{4 \pi}$ and $\gamma=5/3$. 

To fully capture the 3D velocity and magnetic fields, the test has been simulated in 3D with 800$\times$12$\times$12 particles on the left state and 500$\times$12$\times$12 particles on the right state arranged on close-packed triangular lattices.  Results at $t=0.2$ are presented in Fig.~\ref{fig:shock2a} with all shock features, with the red line giving the solution from \citetalias{rj95}.  No post-shock noise in the magnetic field is evident, indicating that the applied artificial resistivity is sufficient.  The L1 error in the $B_y$ profile is $3.086\times 10^{-3}$, compared to $3.358\times 10^{-3}$ if the \citetalias{pm05} switch is used instead,  and for the $B_z$ profile is $5.33\times 10^{-3}$ for our new switch and $6.203\times 10^{-3}$ for the \citetalias{pm05} switch. 

\subsection{Shocktube 5A}
\label{sec:shock5a}

\begin{figure}
 \includegraphics[width=1.0\linewidth]{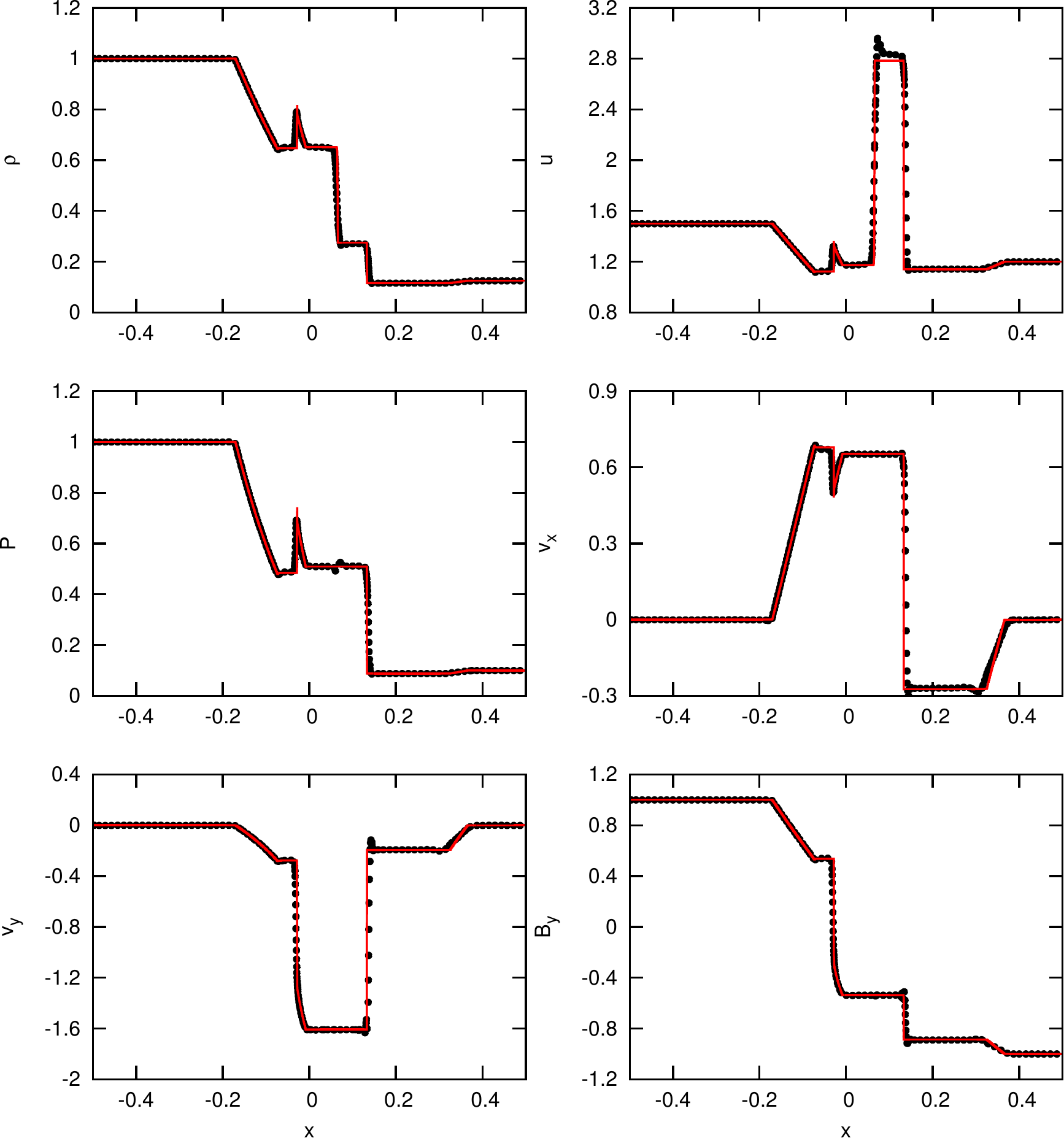}
\caption{Shocktube test 5A from \citetalias{rj95} performed in 2D with left state ($\rho$, $P$, $v_x$, $v_y$, $B_y$) $=$ (1, 1, 0, 0, 1) and right state ($\rho$, P, $v_x$, $v_y$, $B_y$) $=$ (0.125, 0.1, 0, 0, -1) with $B_x=0.75$ at $t=0.1$. Black circles are the particles and the red line is the solution obtained with the Athena code using $10^4$ grid cells.}
\label{fig:shock5a}
\end{figure}

The final shocktube originates from \citet{briowu88}.  It is another 2D shocktube, however it is of particular interest as it contains a compound shock/rarefaction structure.  It has the same initial density and pressure profile as the standard Sod shocktube \citep{sod78}, but with the addition of a magnetic field.  The left state is ($\rho$, $P$, $v_x$, $v_y$, $B_y$) $=$ (1, 1, 0, 0, 1) and right state ($\rho$, P, $v_x$, $v_y$, $B_y$) $=$ (0.125, 0.1, 0, 0, -1) with $B_x=0.75$.  Here we use $\gamma=5/3$ instead of $2$ to follow the results of \citetalias{rj95}.  

The shock has been simulated with 800$\times$30 particles for the right state and 300$\times$10 particles for the right state.  Results at $t=0.1$ are presented in Fig.~\ref{fig:shock5a}.  For this test, the Riemann solution of \citetalias{rj95} does not contain the slow compound structure, so instead we compare our results against those from the Athena code \citep{athena} using $10^4$ grid cells.  As previously, no post-shock noise in the magnetic field is found.  The L1 error profile for $B_y$ is $4.231\times 10^{-3}$ when using our new switch, compared to $6.259 \times 10^{-3}$ if the \citetalias{pm05} switch is used.

\subsection{Polarised Alfv\'en Wave}
\label{sec:polarizedalfven}

We now examine the ability of the switch to reduce dissipation when no shocks are present.  The test problem used is a circularly polarised Alfv\'en wave travelling in a 2D periodic box, following \citet{toth2000}.  This is an exact solution to the ideal MHD equations, so the wave should return to its original state after each crossing.  There are no discontinuities in the magnetic field in this test, but gradients in the magnetic field may cause the $\alpha_\text{B}$ switch to activate.

The simulation is set up using $1682$ particles arranged on a triangular lattice in a periodic domain of lengths $[x,y] = [1/\cos(\omega), 1/\sin(\omega)]$ using $\omega=\pi/6$ which sets the direction of wave motion.  The initial density and pressure are $\rho=1$ and $P=0.1$ with $\gamma=5/3$.  The velocity and magnetic fields parallel and perpendicular to the wave are $[v_\parallel, v_\perp] = [0, 0.1 \sin(2 \pi x_\xi)]$, and $[B_\parallel, B_\perp] = [1, 0.1 \sin(2 \pi x_\xi)]$ where $x_\xi = x \cos(\omega) + y \sin(\omega)$.  Velocity and magnetic field components oriented out of the plane are $v_z = B_z = 0.1 \cos(2 \pi x_\xi)$.

\begin{figure}
 \includegraphics[width=\linewidth]{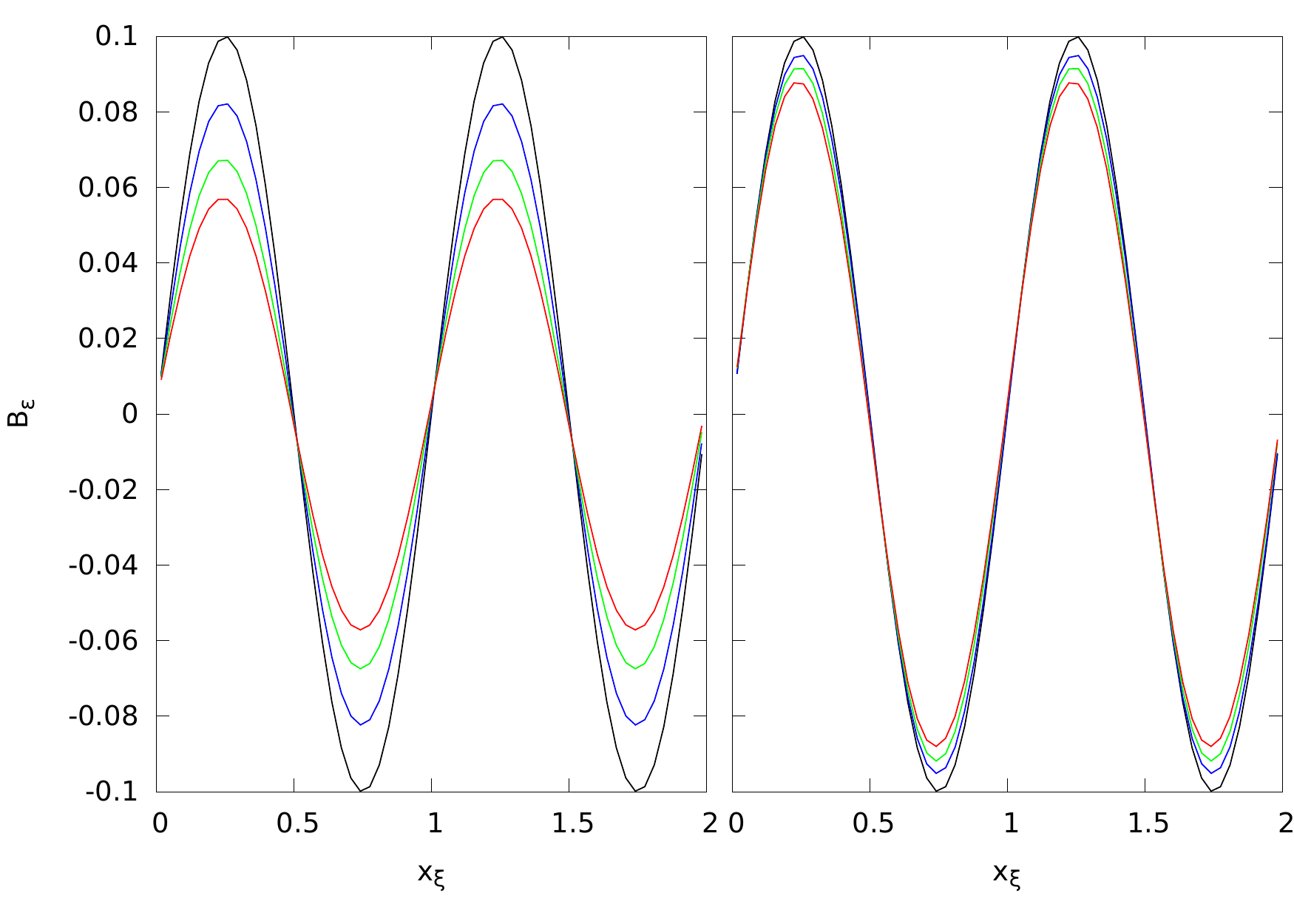}
\caption{Results of the polarised Alfv\'en wave propagation test in 2D, with the exact solution in black, and at $t=2,4,6$ corresponding to 2, 4, and 6 periods.  On the left, the \citetalias{pm05} switch has been used whereas on the right the new resistivity switch has been used.  The maximum $\alpha_\text{B}$ values are $10\times$ higher for the \citetalias{pm05} switch than the new switch, and after 6 periods the amplitude of the wave has decayed over $40\%$ for the \citetalias{pm05} switch compared to only $10\%$ for the new switch.}
\label{fig:polarizedalfven}
\end{figure}

The value of $\alpha_\text{B}$ produced using the new switch can be calculated from the initial conditions, which give $\vert \nabla {\bf B} \vert = 0.2 \pi$ and $\vert {\bf B} \vert = 1$. Thus, for a smoothing length $h=1.2 \Delta x$ where $\Delta x$ is the particle spacing, the new switch gives $\alpha_\text{B}\sim 0.02$ at this resolution. By contrast, the simulations using the \citetalias{pm05} switch produce maximum $\alpha_\text{B}$ values approximately $10 \times$ higher ($0.22$ vs $0.02$), meaning that in this case the \citetalias{pm05} switch is an order of magnitude more dissipative at $t=0$.

After 6 periods, the amplitude of the wave has decayed by over $40\%$ using the \citetalias{pm05} compared to only $\sim 10\%$ for the new switch.  Although the maximum $\alpha_\text{B}$ is $10 \times$ higher with the \citetalias{pm05} switch than the new switch, this is not reflected in the wave amplitude after 6 periods because $\vert \nabla {\bf B} \vert$ and the source term in Eq.~\ref{eq:intalphab} are reduced as the wave is damped.  The rate of this reduction differs between the two switches since the \citetalias{pm05} switch damps the wave more heavily.

\subsection{Orszag-Tang vortex}
\label{sec:orszag}

The Orszag-Tang vortex \citep{ot79} is a widely used test for many astrophysical MHD codes \citep[e.g.,][]{athena, ramses, mhdgadget}.   The problem has an initial vortex structure creating several classes of interacting shock waves which evolve into turbulence.  The initial structure has $\rho=25/(36\pi)$, $P=5/(12\pi)$, ${\bf v}=[-\sin(2\pi y), \sin(2\pi x)]$, and ${\bf B}=[-\sin(2\pi y), \sin(4\pi x)]$ with $\gamma=5/3$. 

The test has been simulated using $512^2$, $1024^2$, and $2048^2$ particles initially arranged on a square lattice.  The initial conditions are set up by first creating the particles in one quadrant of the domain, then mirroring the configuration to the other quadrants with appropriate sign changes in the velocity and magnetic fields as needed.  This removes the slight discrepancies from floating point arithmetic, retaining exact symmetry in the initial conditions.

% rho = [0.08, 0.35]
% Bpres = [0, 0.18]
% alpha = [0, 1]

\begin{figure}
 \setlength{\tabcolsep}{0.002\textwidth}
\begin{tabular}{ccl}
 \scriptsize{PM05 switch} & \scriptsize{New switch} 
\\
   \includegraphics[height=0.445\linewidth]{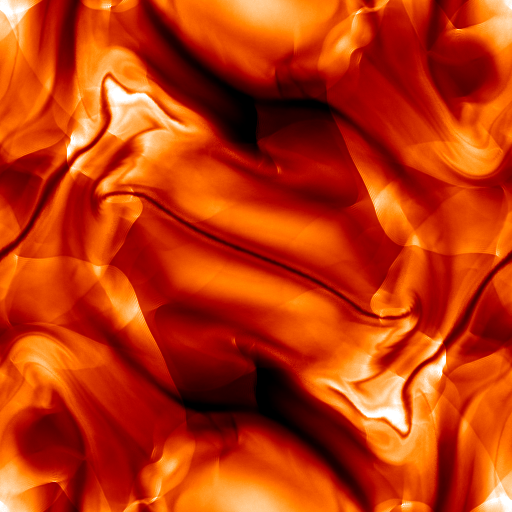} 
 & \includegraphics[height=0.445\linewidth]{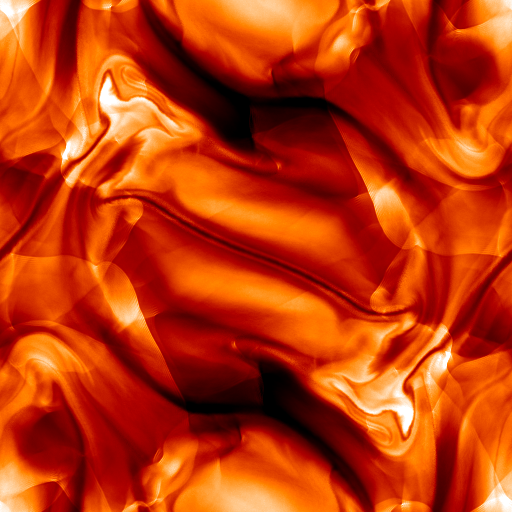} 
 & \includegraphics[height=0.445\linewidth]{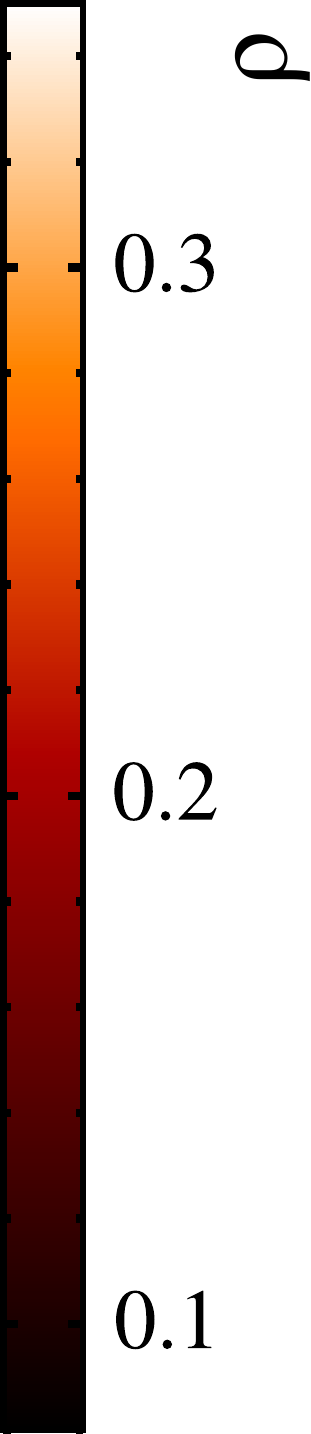}
\\
   \includegraphics[height=0.445\linewidth]{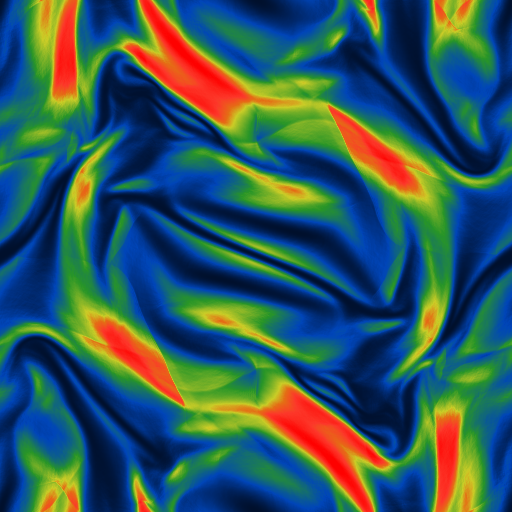}
 & \includegraphics[height=0.445\linewidth]{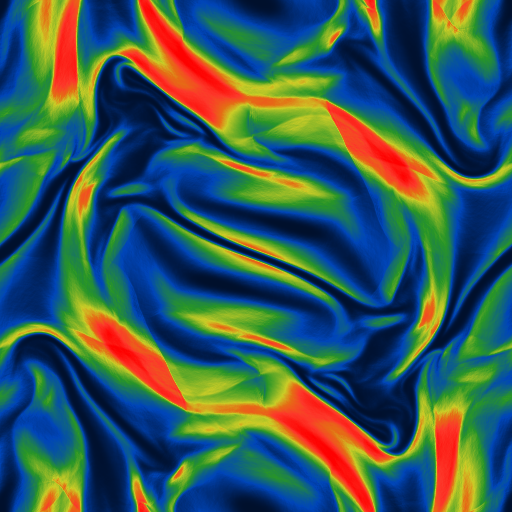}
 & \includegraphics[height=0.445\linewidth]{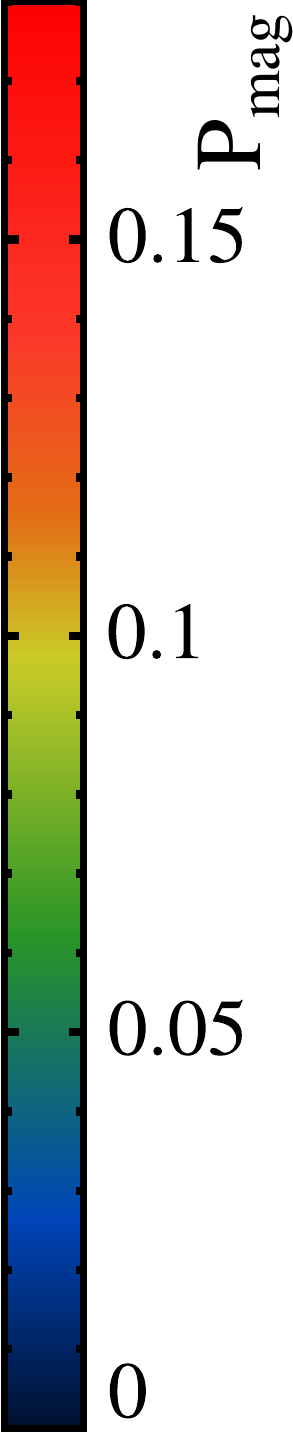}
\\
   \includegraphics[height=0.445\linewidth]{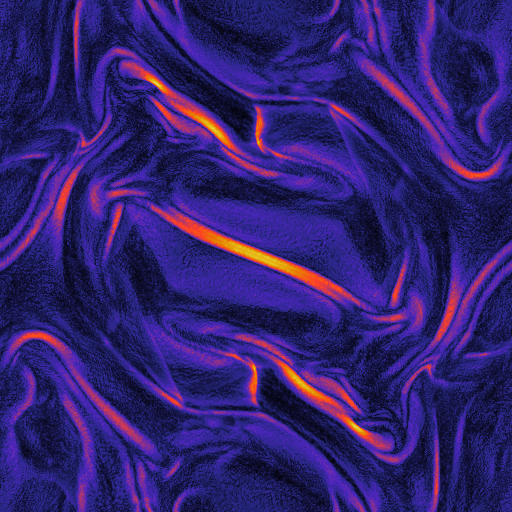}
 & \includegraphics[height=0.445\linewidth]{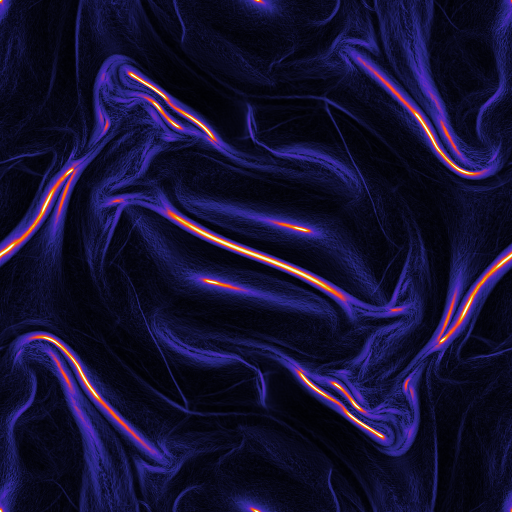}
 & \includegraphics[height=0.445\linewidth]{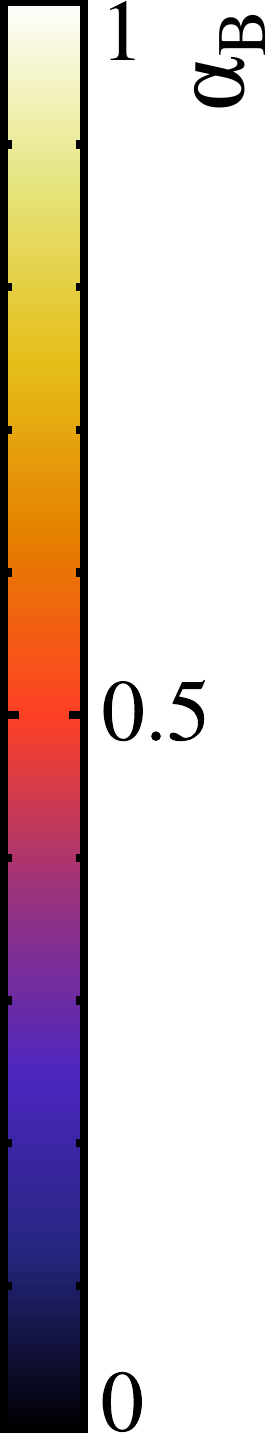}
\\
\end{tabular}
\caption{The density (top), magnetic pressure (middle), and $\alpha_\text{B}$ (bottom) of the Orszag-Tang vortex at $t=1$ for the old (left) and new (right) resistivity switches.  The new switch effectively traces the shock lines, with little or no dissipation between shocks.  The low density regions are more sharply defined using the new switch due to the decreased dissipation of the magnetic field structure.}
\label{fig:orszag-alphab}
\end{figure}

% @512: alphaB mean = 0.205 old
% @512: alphaB mean = 0.110 new

% @1024: alphaB mean = 0.145 old
% @1024: alphaB mean = 0.070 new

% @2048: alphaB mean = 0.097 old
% @2048: alphaB mean = 0.046 new

Results are presented at $t=1$ in Fig.~\ref{fig:orszag-alphab} which shows renderings of the density, magnetic pressure, and $\alpha_\text{B}$ in the domain for $1024^2$ particles.  The new switch is effective at activating resistivity along the shock lines, yet keeps $\alpha_\text{B}$ minimal between shocks.  By contrast, the \citetalias{pm05} switch results in broad regions with $\alpha_\text{B}\approx 1$ near shocks and a mean $\alpha_\text{B}$ twice as high ($\sim0.2$ to $\sim0.1$).  This leads to a smoothing away of subtle magnetic features, particularly noticeable around the central magnetic feature, and in some of the low density regions which are less sharply defined.

\begin{figure}
 \includegraphics[width=\linewidth]{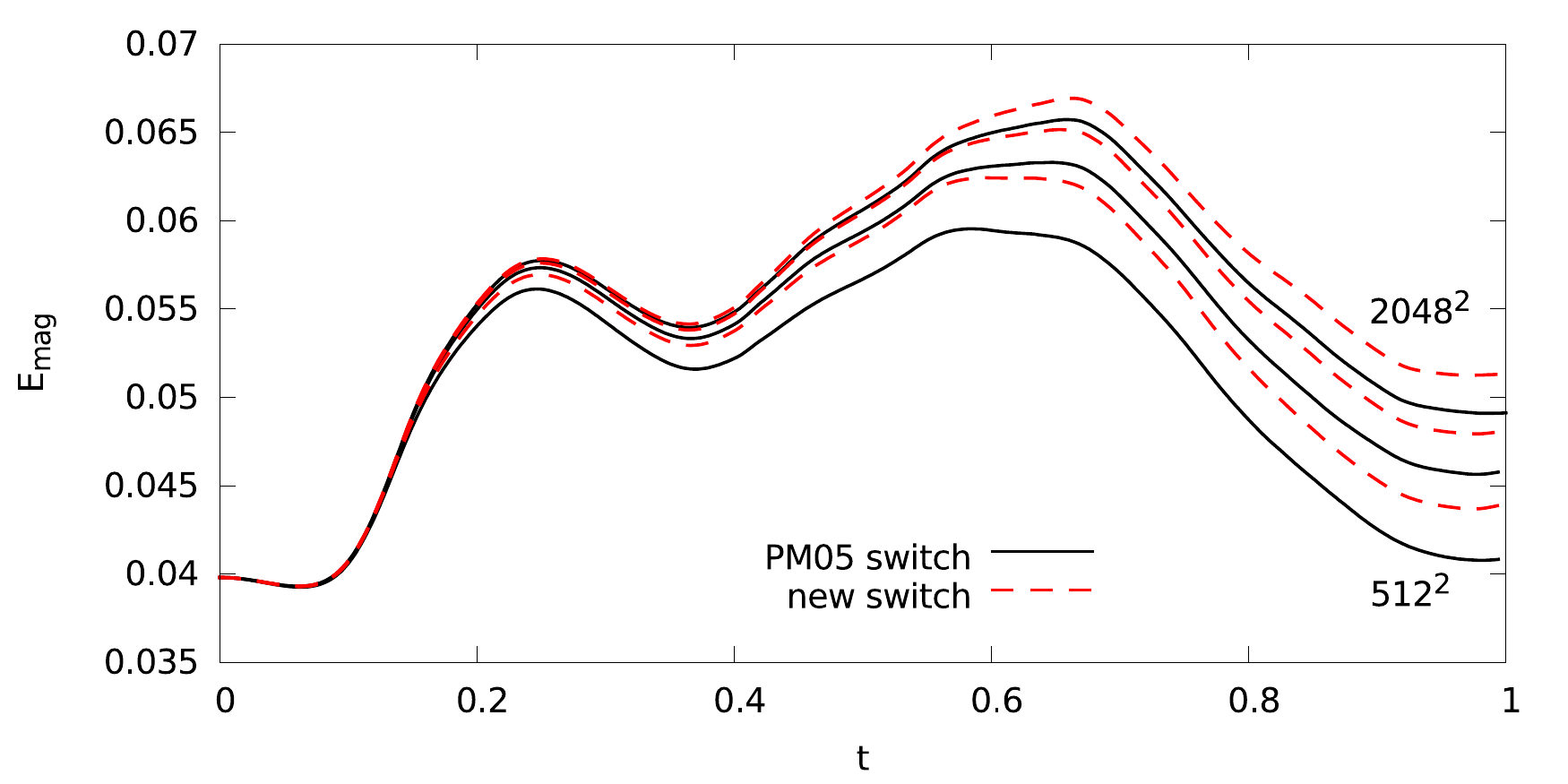}
\caption{Evolution of the magnetic energy for the Orszag-Tang vortex using the PM05 resistivity switch (black, solid lines) and the new resistivity switch (red, dashed lines) at resolutions of $512^2$, $1024^2$, and $2048^2$ particles.  The new switch is much less dissipative than the PM05 switch, producing an effect similar to increasing the resolution.}
\label{fig:orszag-be}
\end{figure}

Fig.~\ref{fig:orszag-be} shows the evolution of the magnetic energy as a function of time for $512^2$, $1024^2$, and $2048^2$ particles.  This shows that the magnetic energy is dissipated less at higher resolution.  Using the new artificial resistivity switch also leads to a lower dissipation rate compared to the \citetalias{pm05} switch, producing an effect equivalent to running the test at higher resolution.

\subsection{Mach 10 MHD turbulence}
\label{sec:mhdturb}

Our final test is of supersonic magnetised turbulence which is representative of conditions in molecular clouds \citep[see reviews by][]{evans99,es04,mo07}.  A stochastic, solenoidal driving force is applied, generating turbulence with a root-mean-square Mach number of 10.  It has an initially weak magnetic field, with the kinetic energy approximately 10 orders of magnitude larger than magnetic energy, which grows through dynamo amplification by the conversion of kinetic to magnetic energy \citep[see review by][]{bs05}.  Our simulations follow the SPH Mach 10 turbulence study of \citet{pricefederrath10}, but in the MHD case of turbulent dynamo amplification studied by \citet{federrathetal11}.

The simulation is set up at a resolution of $128^3$ particles.  The initial density is $\rho=1$ with an isothermal equation of state using a speed of sound of $c=1$.  The gas is initially at rest, and has a uniform magnetic field $B_z = \sqrt{2}\times 10^{-5}$ such that the initial plasma $\beta = 10^{10}$.  

To drive the turbulence, an acceleration based on an Ornstein-Uhlenbeck process is used \citep{eswaranpope88, federrathetal10}, which is a stochastic process with a finite autocorrelation timescale that drives motion at low wave numbers.  The driving force is constructed in Fourier space, allowing it to be decomposed into solenoidal and compressive components and for this case we only use the solenoidal component.

\begin{figure}
 %\centering
\setlength{\tabcolsep}{0.002\textwidth}
\begin{tabular}{cccr}
\scriptsize{Fixed $\alpha_\text{B}=1$} & \scriptsize{PM05 switch} & \scriptsize{New switch} 
\\
   \includegraphics[height=0.30\linewidth]{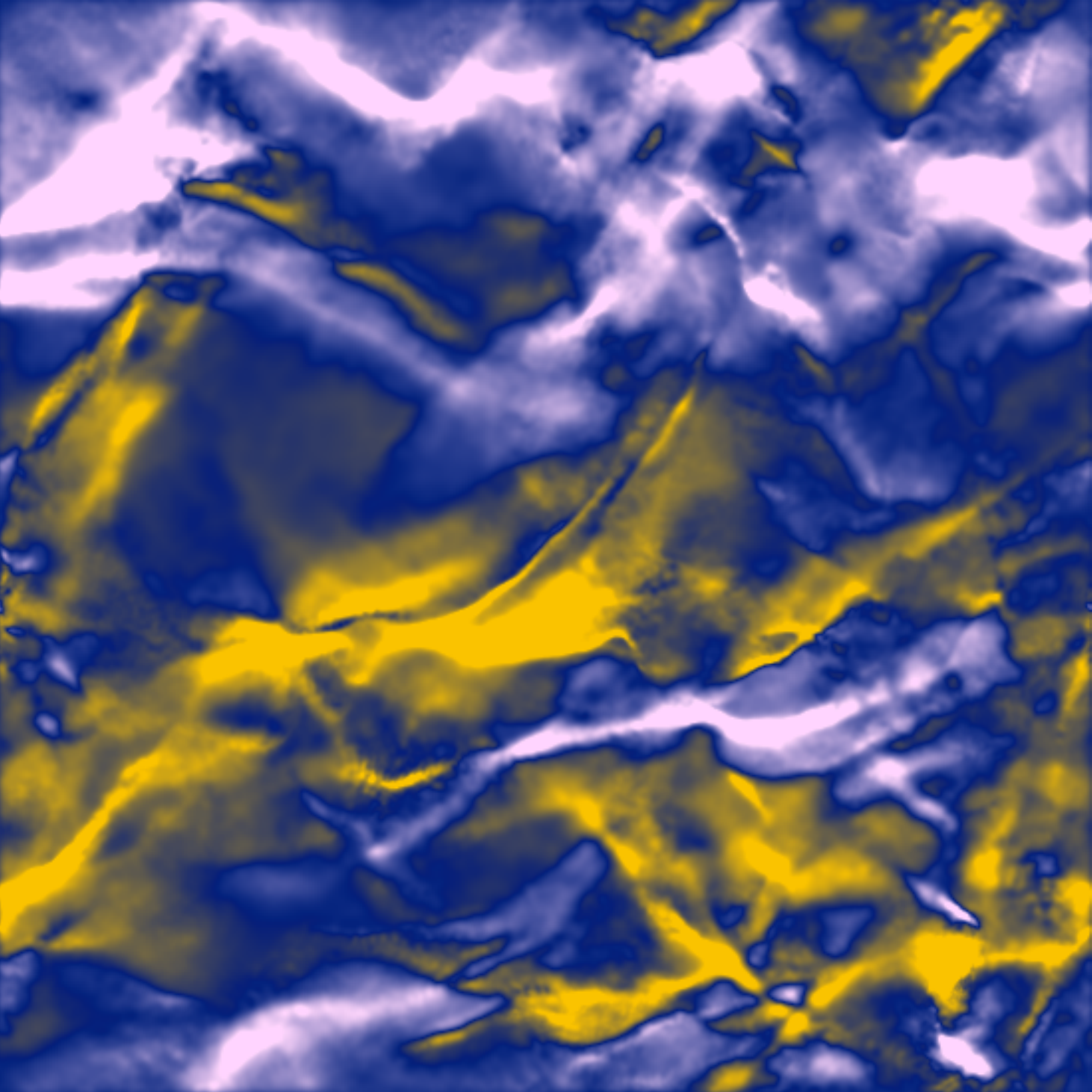}
 & \includegraphics[height=0.30\linewidth]{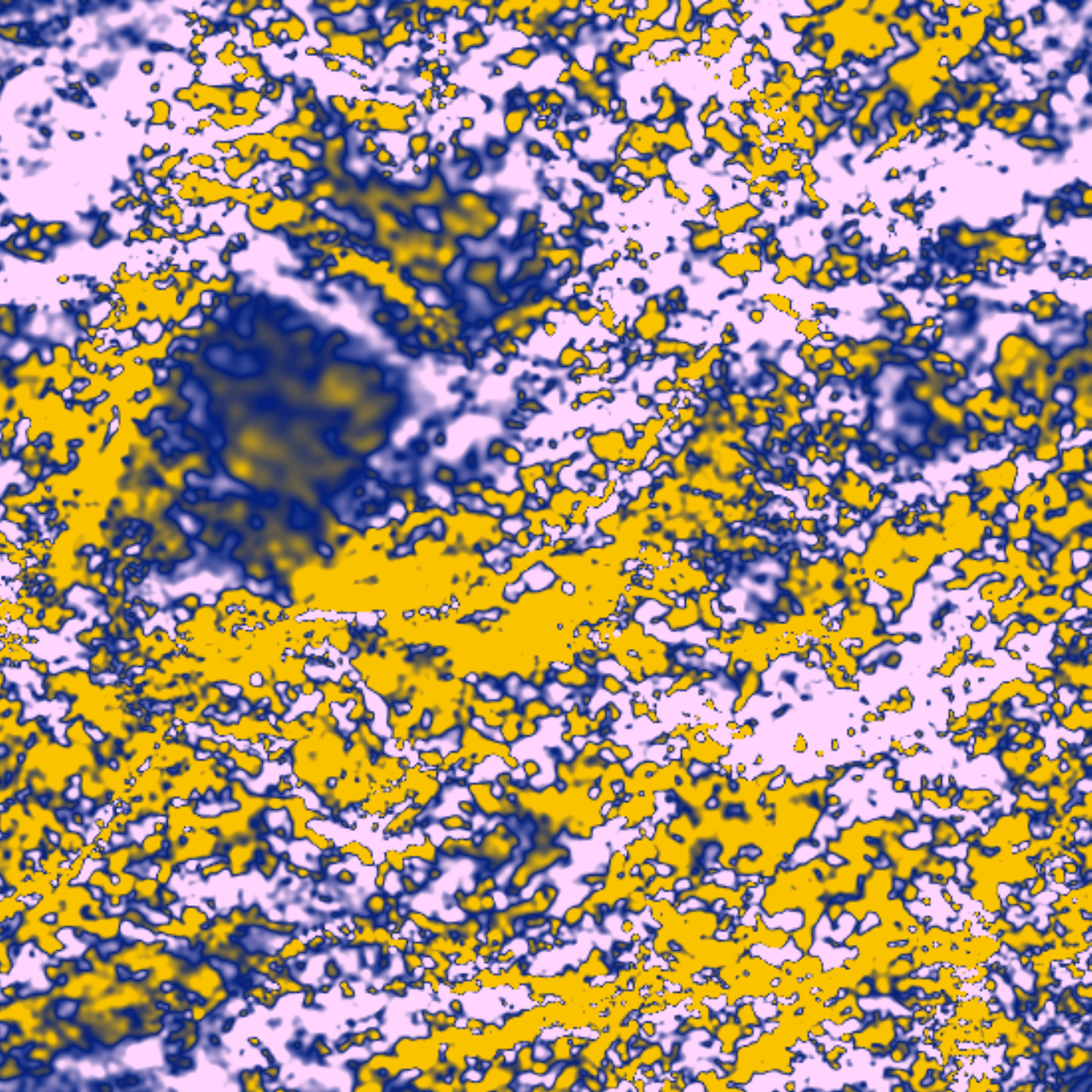}
 & \includegraphics[height=0.30\linewidth]{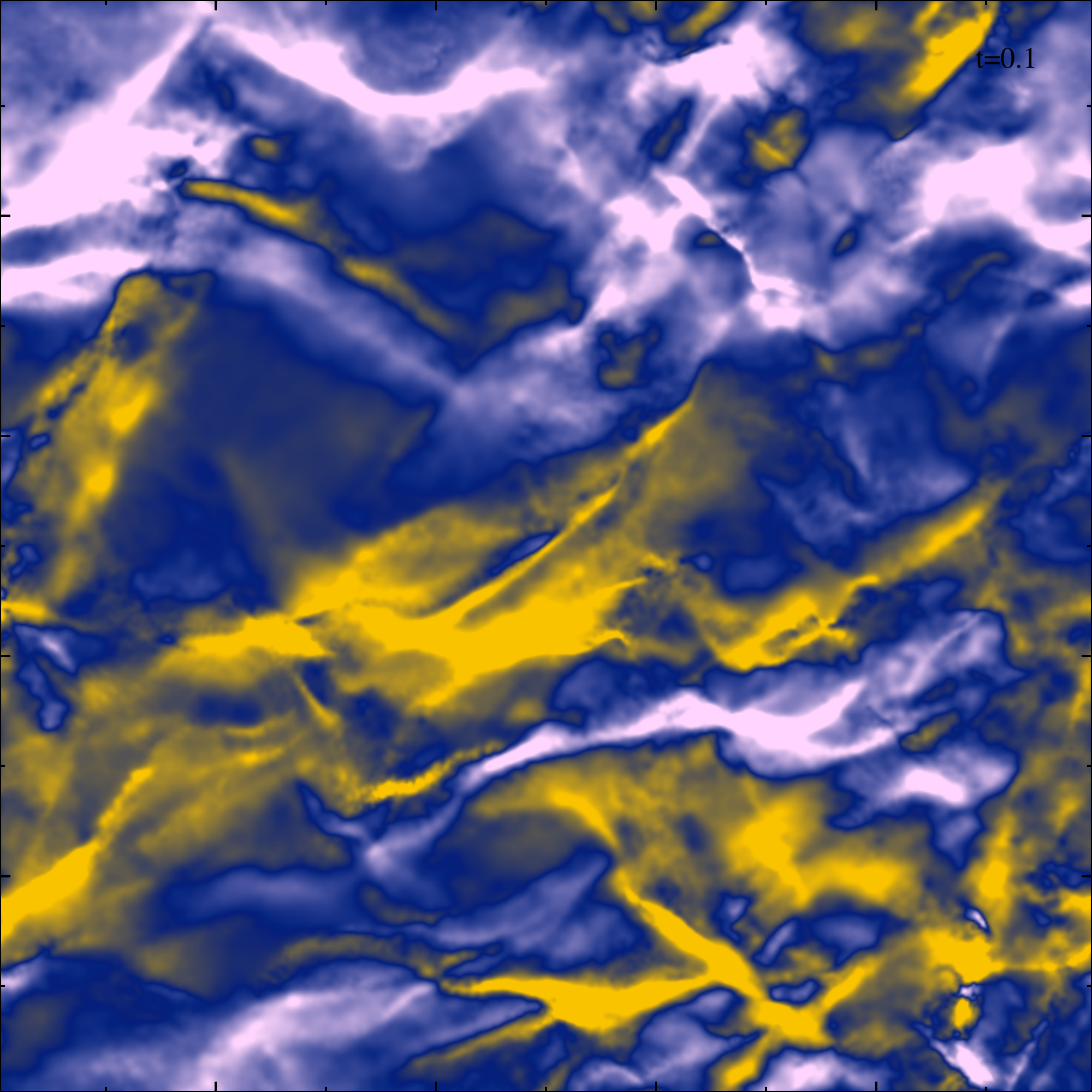}
 & \includegraphics[height=0.30\linewidth]{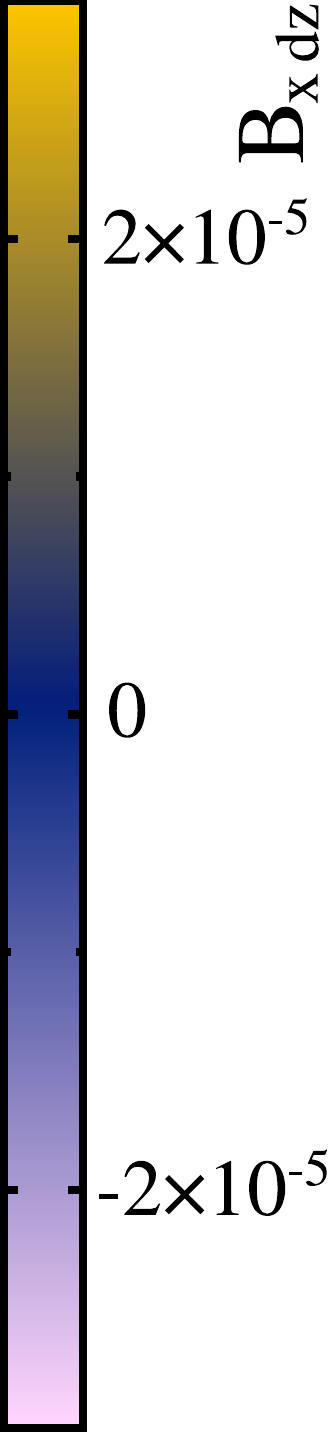}
\\
   \includegraphics[height=0.30\linewidth]{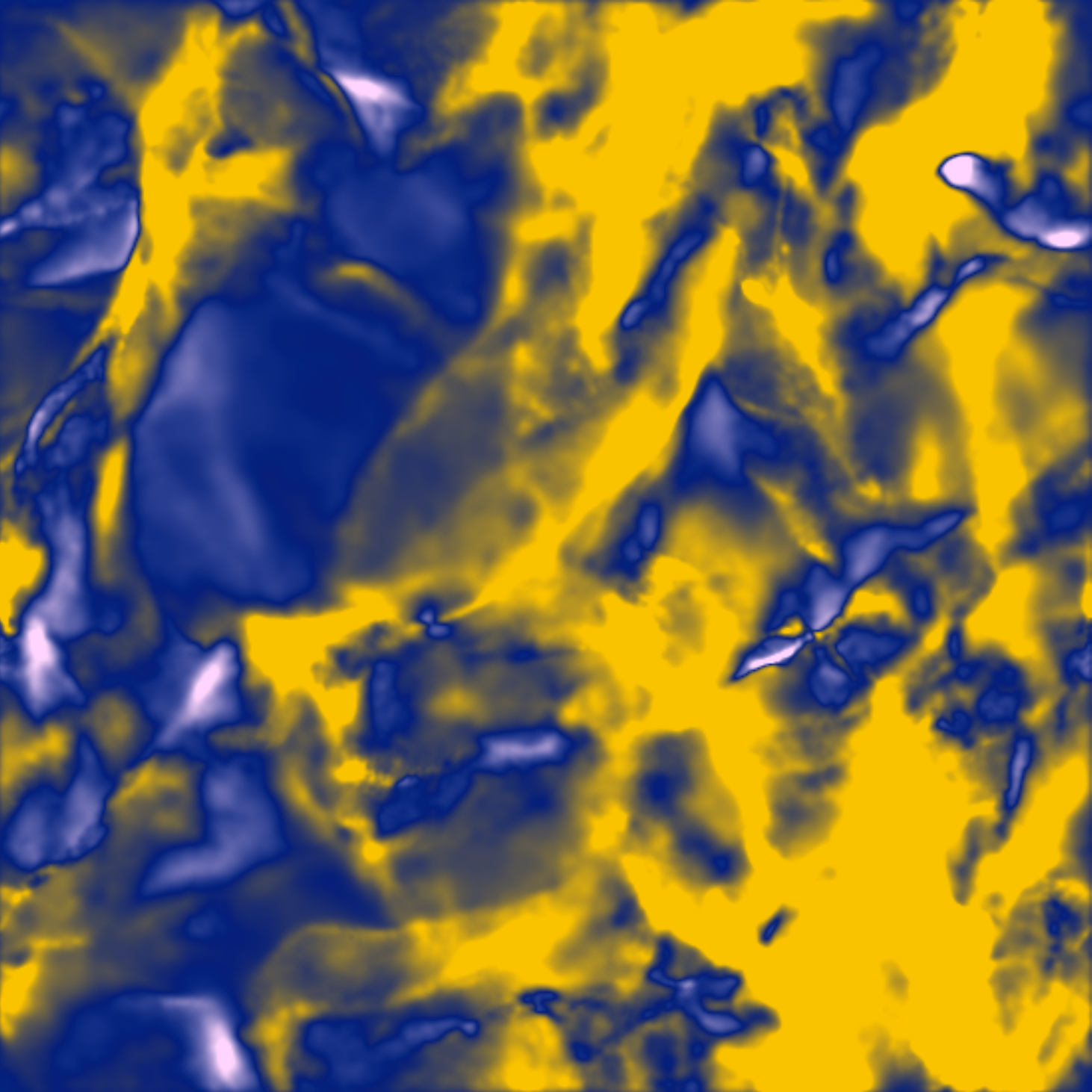}
 & \includegraphics[height=0.30\linewidth]{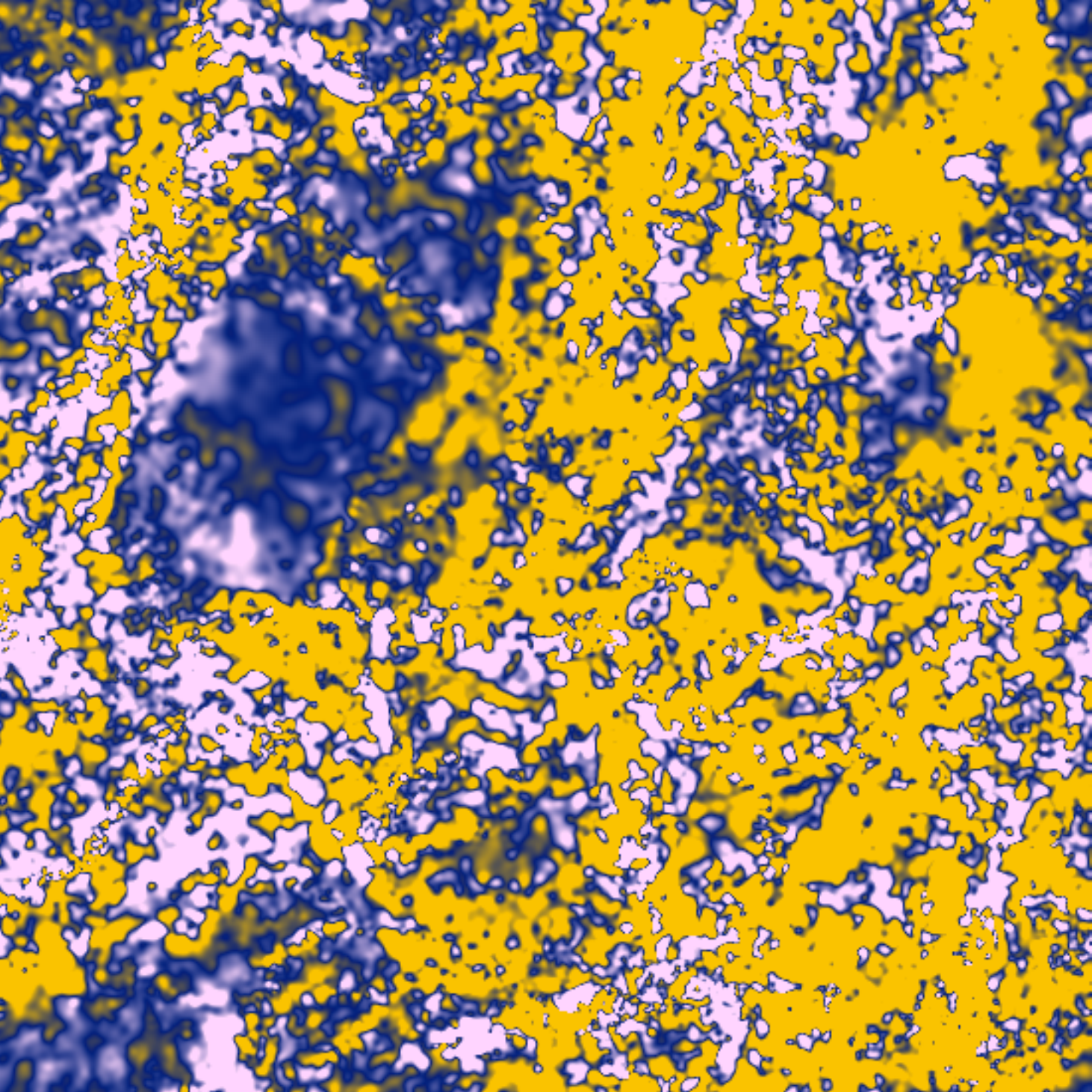}
 & \includegraphics[height=0.30\linewidth]{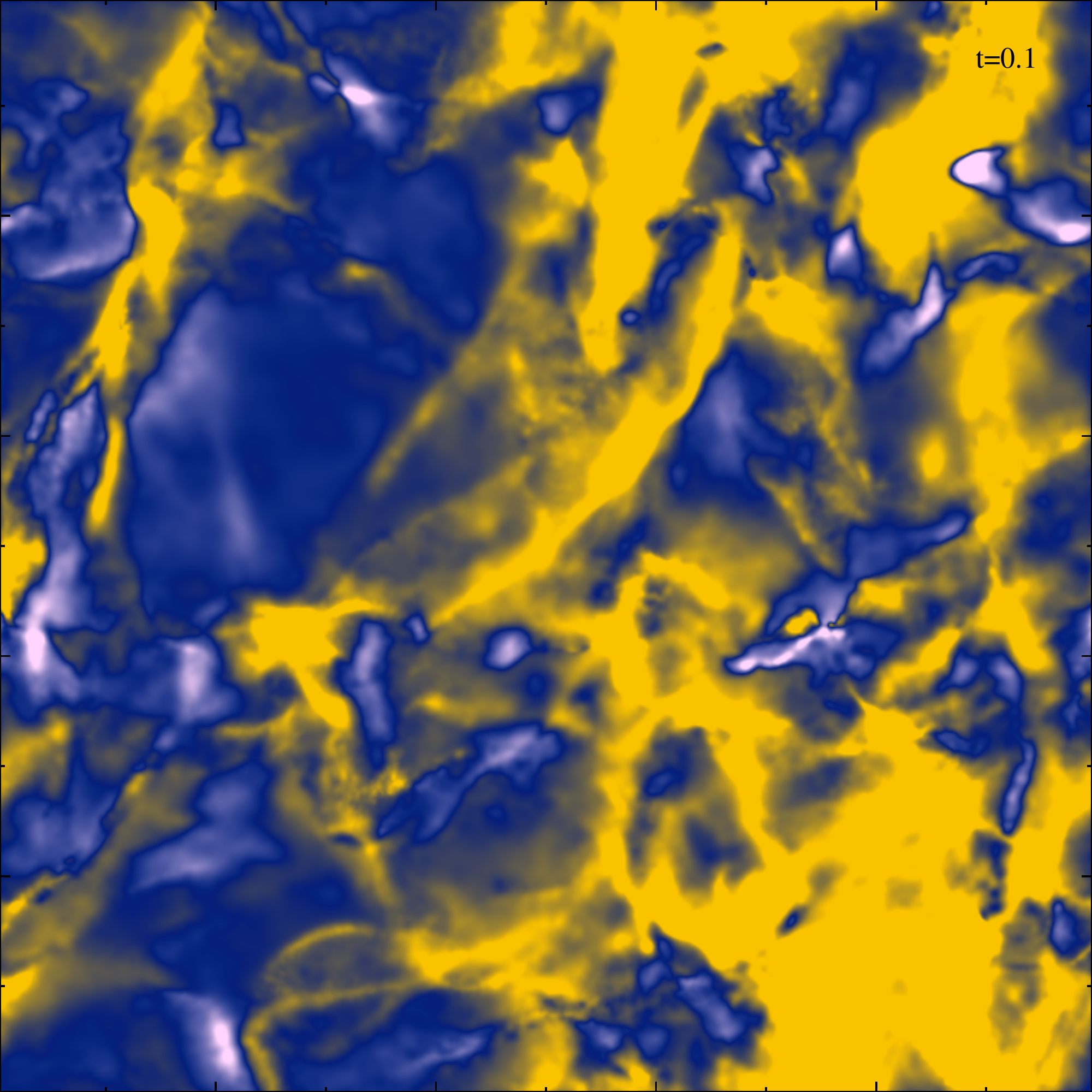}
 & \includegraphics[height=0.30\linewidth]{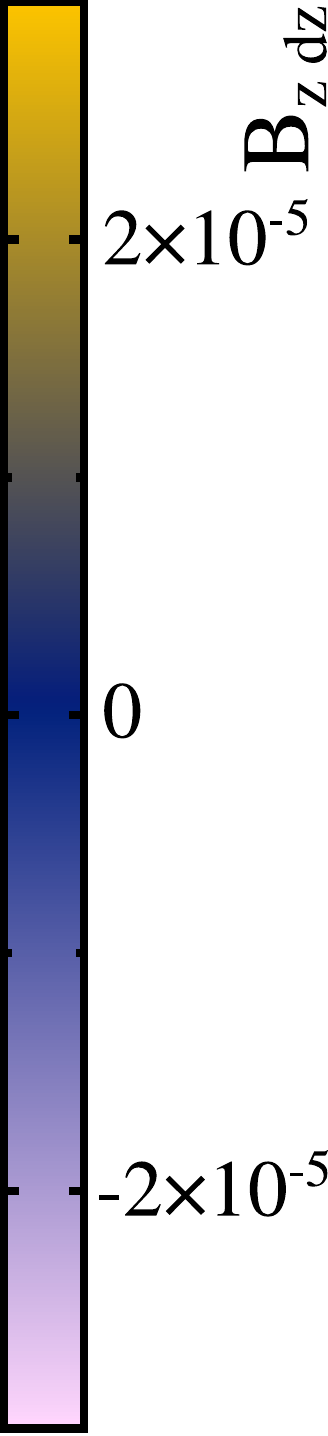}
\end{tabular}
\caption{The column integrated $x$ \& $z$ (top, bottom) magnetic field components using fixed $\alpha_\text{B}=1$ (left), the \citetalias{pm05} switch (centre), and the new switch (right) after two turbulent turnover times (i.e., the regime of fully developed turbulence).  The magnetic field structure using the previous switch is dominated by unphysical noise due to the shocks failing to be captured (centre), whereas the new switch is able to capture the shocks and the magnetic field retains its physical structure (right).}
\label{fig:mhdturb}
\end{figure}

The column integrated $x$ \& $z$ components of the magnetic field are shown in Fig.~\ref{fig:mhdturb} at $t=2$ turbulent turnover times.  The \citetalias{pm05} switch fails to raise $\alpha_\text{B}$ to appreciable levels ($\alpha_\text{B} \sim 10^{-5}$), and as demonstrated in Fig.~\ref{fig:mhdturb}, the shocks in the magnetic field fail to be captured.  This leads to break-up of the shocks, causing unphysical magnetic field growth until such a time as the field is strong enough to activate the switch.  By contrast, the new switch is invariant to field strength meaning that it turns on resistivity in the shocked regions and the shocks are captured.

We also found for this simulation that using the averaged Alfv\'en speed as the signal velocity for resistivity produces the same behaviour (shocks breaking apart).  In this instance, it is due to the large disparity between the Alfv\'en and sound speed meaning that the applied resistivity is too weak to capture the strong shocks properly.  With the fast MHD wave speed in the signal velocity (Eq.~\ref{eq:vsigfastmhd}), the shocks are captured correctly.

\section{Summary}
\label{sec:conclusion}
We have developed a switch to dynamically regulate the amount of artificial resistivity applied to the magnetic field in Smoothed Particle Magnetohydrodynamics simulations. Since the purpose of artificial resistivity is to model magnetic shocks and discontinuities, the key is to minimise spurious dissipation in smooth parts of the field.  Our switch accomplishes this by setting the artificial resistivity parameter $\alpha_\text{B}$ equal to the dimensionless quantity $h \vert \nabla {\bf B} \vert / \vert {\bf B} \vert$. This yields a simple, powerful and robust method for reducing magnetic dissipation away from shocks with no loss in shock-capturing ability. Importantly, it responds appropriately at all magnetic field strengths, a particular improvement over the PM05 switch which was found to inadequately capture shocks in weak fields.

Alternative switches using the second derivative of the magnetic field were also investigated, in particular $h \vert \nabla^2 {\bf B} \vert / \vert \nabla {\bf B} \vert$ and $h^2 \vert \nabla^2 {\bf B} \vert / \vert {\bf B} \vert$.  The key requirement to their success is a high order estimate of the second derivative, otherwise noise from particle disorder overwhelms the derivative estimate and causes excessive dissipation.  Obtaining this higher order estimate, however, adds significant computational expense.  In general, we recommend our first derivative switch for normal use since it is simple, yet performs robustly and effectively.

Three shocktube tests (Sections~\ref{sec:shock1b}, \ref{sec:shock2a}, and \ref{sec:shock5a}) were used to establish that the switch correctly models a range of shock phenomena, including fast and slow shocks, fast and slow rarefactions, rotational discontinuities, and compound shock structures. The L1 error of the magnetic field profiles for all tests was lower when using this new switch compared to using the \citetalias{pm05} switch. These tests also demonstrated that the \citet{tricco&price12} divergence cleaning algorithm is stable and robust in shock tube problems, in contrast to the version proposed by \citet{sdb12}.

In Section~\ref{sec:polarizedalfven}, the propagation of a travelling Alfv\'en wave was used to gauge the switch's ability to reduce unwanted dissipation in situations not involving discontinuities, and was found to result in maximum $\alpha_\text{B}$ values 10$\times$ smaller than the \citetalias{pm05} switch ($\sim0.02$ compared to $\sim0.22$). After 6 periods, the amplitude of the wave using the \citetalias{pm05} switch was four times lower than using the new switch.
 
The Orszag-Tang vortex was used in Section~\ref{sec:orszag} to examine the performance of the new switch when there are multiple interacting shocks, producing regions of $\alpha_\text{B}\sim 1$ that closely traced the shock lines. The new switch was found to decrease the spurious dissipation in smooth regions compared to the \citetalias{pm05} switch, leading to the subtle magnetic features being more sharply defined, equivalent to running the test at higher resolution.

Finally, in Section~\ref{sec:mhdturb}, a simulation of Mach 10 MHD turbulence was used to demonstrate the switch's ability to capture magnetic shocks when a weak magnetic field is combined with strong hydrodynamic shocks.  The \citetalias{pm05} switch was found to fail for the low field strengths present in this problem, causing the magnetic field to be dominated by unphysical noise. With the new switch the magnetic shocks remain coherent.

We found that it is very important to use the fast MHD wave speed as the characteristic signal velocity for artificial resistivity. Using the Alfv\'en speed as the characteristic signal velocity, as proposed by \citet{price12}, was found to inadequately capture fast MHD shocks in the highly super-Alfv\'enic regime, leading to unphysical effects (Fig.~\ref{fig:mhdturb}).

Our new switch is widely applicable to astrophysical SPMHD simulations, in particular for simulations involving weak fields such as in galaxy and cosmological simulations, and also for dynamo processes. In every case we tested, it produced lower magnetic dissipation than the \citetalias{pm05} switch, making it possible to achieve higher magnetic Reynolds numbers in simulations of the interstellar and intergalactic medium. The new switch thus supercedes the \citetalias{pm05} in every respect.

\vspace{-0.5cm}
\section{Acknowledgments}

We thank the referee, Walter Dehnen, for his helpful comments and suggestions which have improved the quality of the paper, and Christoph Federrath and Guillaume Laibe for useful discussions.  T. Tricco is supported by Endeavour IPRS and APA postgraduate scholarships.  We are grateful for funding via Australian Research Council Discovery Projects grant DP1094585.  This research was undertaken with the assistance of resources provided at the Multi-modal Australian ScienceS Imaging and Visualisation Environment (MASSIVE) through the National Computational Merit Allocation Scheme supported by the Australian Government.

\bibliographystyle{mn2e}
\bibliography{resisbib}

\begin{thebibliography}{}

\bibitem[\protect\citeauthoryear{{Balsara}}{{Balsara}}{1995}]{balsara95}
{Balsara} D.~S.,  1995, J. Comput. Phys., 121, 357

\bibitem[\protect\citeauthoryear{{B{\o}rve}, {Omang} \& {Trulsen}}{{B{\o}rve}
  et~al.}{2001}]{borve01}
{B{\o}rve} S.,  {Omang} M.,    {Trulsen} J.,  2001, ApJ, 561, 82

\bibitem[\protect\citeauthoryear{{Brandenburg} \& {Subramanian}}{{Brandenburg}
  \& {Subramanian}}{2005}]{bs05}
{Brandenburg} A.,  {Subramanian} K.,  2005, Phys. Rep., 417, 1

\bibitem[\protect\citeauthoryear{{Brio} \& {Wu}}{{Brio} \&
  {Wu}}{1988}]{briowu88}
{Brio} M.,  {Wu} C.~C.,  1988, J. Comput. Phys., 75, 400

\bibitem[\protect\citeauthoryear{{Brookshaw}}{{Brookshaw}}{1985}]{brookshaw85}
{Brookshaw} L.,  1985, Proceedings of the Astronomical Society of Australia, 6,
  207

\bibitem[\protect\citeauthoryear{{Cullen} \& {Dehnen}}{{Cullen} \&
  {Dehnen}}{2010}]{cd10}
{Cullen} L.,  {Dehnen} W.,  2010, MNRAS, 408, 669

\bibitem[\protect\citeauthoryear{{Dai} \& {Woodward}}{{Dai} \&
  {Woodward}}{1994}]{dw94}
{Dai} W.,  {Woodward} P.~R.,  1994, J. Comput. Phys., 115, 485

\bibitem[\protect\citeauthoryear{{Dedner}, {Kemm}, {Kr{\"o}ner}, {Munz},
  {Schnitzer} \& {Wesenberg}}{{Dedner} et~al.}{2002}]{dedneretal02}
{Dedner} A.,  {Kemm} F.,  {Kr{\"o}ner} D.,  {Munz} C.-D.,  {Schnitzer} T.,
  {Wesenberg} M.,  2002, Journal of Computational Physics, 175, 645

\bibitem[\protect\citeauthoryear{{Dolag} \& {Stasyszyn}}{{Dolag} \&
  {Stasyszyn}}{2009}]{mhdgadget}
{Dolag} K.,  {Stasyszyn} F.,  2009, MNRAS, 398, 1678

\bibitem[\protect\citeauthoryear{{Elmegreen} \& {Scalo}}{{Elmegreen} \&
  {Scalo}}{2004}]{es04}
{Elmegreen} B.~G.,  {Scalo} J.,  2004, ARA\&A, 42, 211

\bibitem[\protect\citeauthoryear{{Eswaran} \& {Pope}}{{Eswaran} \&
  {Pope}}{1988}]{eswaranpope88}
{Eswaran} V.,  {Pope} S.~B.,  1988, Computers and Fluids, 16, 257

\bibitem[\protect\citeauthoryear{{Evans} II}{{Evans}}{1999}]{evans99}
{Evans} II N.~J.,  1999, ARA\&A, 37, 311

\bibitem[\protect\citeauthoryear{{Federrath}, {Chabrier}, {Schober},
  {Banerjee}, {Klessen} \& {Schleicher}}{{Federrath}
  et~al.}{2011}]{federrathetal11}
{Federrath} C.,  {Chabrier} G.,  {Schober} J.,  {Banerjee} R.,  {Klessen}
  R.~S.,    {Schleicher} D.~R.~G.,  2011, Phys. Rev. Lett., 107, 114504

\bibitem[\protect\citeauthoryear{{Federrath} \& {Klessen}}{{Federrath} \&
  {Klessen}}{2012}]{fk12}
{Federrath} C.,  {Klessen} R.~S.,  2012, ApJ, 761, 156

\bibitem[\protect\citeauthoryear{{Federrath}, {Roman-Duval}, {Klessen},
  {Schmidt} \& {Mac Low}}{{Federrath} et~al.}{2010}]{federrathetal10}
{Federrath} C.,  {Roman-Duval} J.,  {Klessen} R.~S.,  {Schmidt} W.,    {Mac
  Low} M.-M.,  2010, A\&A, 512, A81

\bibitem[\protect\citeauthoryear{{Fromang}, {Hennebelle} \&
  {Teyssier}}{{Fromang} et~al.}{2006}]{ramses}
{Fromang} S.,  {Hennebelle} P.,    {Teyssier} R.,  2006, Astron. Astrophys.,
  457, 371

\bibitem[\protect\citeauthoryear{{Gaensler}, {Haverkorn}, {Burkhart},
  {Newton-McGee}, {Ekers}, {Lazarian}, {McClure-Griffiths}, {Robishaw},
  {Dickey} \& {Green}}{{Gaensler} et~al.}{2011}]{gaensleretal11}
{Gaensler} B.~M.,  {Haverkorn} M.,  {Burkhart} B.,  {Newton-McGee} K.~J.,
  {Ekers} R.~D.,  {Lazarian} A.,  {McClure-Griffiths} N.~M.,  {Robishaw} T.,
  {Dickey} J.~M.,    {Green} A.~J.,  2011, Nature, 478, 214

\bibitem[\protect\citeauthoryear{{Hubber}, {Falle} \& {Goodwin}}{{Hubber}
  et~al.}{2013}]{hubberetal13}
{Hubber} D.~A.,  {Falle} S.~A.~E.~G.,    {Goodwin} S.~P.,  2013, MNRAS, 432,
  711

\bibitem[\protect\citeauthoryear{{McKee} \& {Ostriker}}{{McKee} \&
  {Ostriker}}{2007}]{mo07}
{McKee} C.~F.,  {Ostriker} E.~C.,  2007, ARA\&A, 45, 565

\bibitem[\protect\citeauthoryear{{Monaghan}}{{Monaghan}}{1997}]{monaghan97}
{Monaghan} J.~J.,  1997, J. Comput. Phys., 136, 298

\bibitem[\protect\citeauthoryear{{Monaghan}}{{Monaghan}}{2005}]{monaghan05}
{Monaghan} J.~J.,  2005, Rep. Prog. Phys., 68, 1703

\bibitem[\protect\citeauthoryear{{Monaghan} \& {Gingold}}{{Monaghan} \&
  {Gingold}}{1983}]{mg83}
{Monaghan} J.~J.,  {Gingold} R.~A.,  1983, Journal of Computational Physics,
  52, 374

\bibitem[\protect\citeauthoryear{{Morris} \& {Monaghan}}{{Morris} \&
  {Monaghan}}{1997}]{mm97}
{Morris} J.~P.,  {Monaghan} J.~J.,  1997, J. Comput. Phys., 136, 41

\bibitem[\protect\citeauthoryear{{Orszag} \& {Tang}}{{Orszag} \&
  {Tang}}{1979}]{ot79}
{Orszag} S.~A.,  {Tang} C.-M.,  1979, J. Fluid Mech., 90, 129

\bibitem[\protect\citeauthoryear{{Phillips} \& {Monaghan}}{{Phillips} \&
  {Monaghan}}{1985}]{phillipsmonaghan85}
{Phillips} G.~J.,  {Monaghan} J.~J.,  1985, MNRAS, 216, 883

\bibitem[\protect\citeauthoryear{{Price}}{{Price}}{2008}]{price08}
{Price} D.~J.,  2008, J. Comput. Phys., 227, 10040

\bibitem[\protect\citeauthoryear{{Price}}{{Price}}{2010}]{price10}
{Price} D.~J.,  2010, MNRAS, 401, 1475

\bibitem[\protect\citeauthoryear{{Price}}{{Price}}{2012}]{price12}
{Price} D.~J.,  2012, J. Comput. Phys., 231, 759

\bibitem[\protect\citeauthoryear{{Price} \& {Bate}}{{Price} \&
  {Bate}}{2007}]{pricebate07}
{Price} D.~J.,  {Bate} M.~R.,  2007, MNRAS, 377, 77

\bibitem[\protect\citeauthoryear{{Price} \& {Federrath}}{{Price} \&
  {Federrath}}{2010}]{pricefederrath10}
{Price} D.~J.,  {Federrath} C.,  2010, MNRAS, 406, 1659

\bibitem[\protect\citeauthoryear{{Price} \& {Monaghan}}{{Price} \&
  {Monaghan}}{2004a}]{pm04a}
{Price} D.~J.,  {Monaghan} J.~J.,  2004a, MNRAS, 348, 123

\bibitem[\protect\citeauthoryear{{Price} \& {Monaghan}}{{Price} \&
  {Monaghan}}{2004b}]{pm04b}
{Price} D.~J.,  {Monaghan} J.~J.,  2004b, MNRAS, 348, 139

\bibitem[\protect\citeauthoryear{{Price} \& {Monaghan}}{{Price} \&
  {Monaghan}}{2005}]{pm05}
{Price} D.~J.,  {Monaghan} J.~J.,  2005, MNRAS, 364, 384

\bibitem[\protect\citeauthoryear{{Price}, {Tricco} \& {Bate}}{{Price}
  et~al.}{2012}]{ptb12}
{Price} D.~J.,  {Tricco} T.~S.,    {Bate} M.~R.,  2012, MNRAS, 423, L45

\bibitem[\protect\citeauthoryear{{Read} \& {Hayfield}}{{Read} \&
  {Hayfield}}{2012}]{rh12}
{Read} J.~I.,  {Hayfield} T.,  2012, MNRAS, 422, 3037

\bibitem[\protect\citeauthoryear{{Rosswog} \& {Price}}{{Rosswog} \&
  {Price}}{2007}]{rosswogprice07}
{Rosswog} S.,  {Price} D.,  2007, MNRAS, 379, 915

\bibitem[\protect\citeauthoryear{{Ryu} \& {Jones}}{{Ryu} \&
  {Jones}}{1995}]{rj95}
{Ryu} D.,  {Jones} T.~W.,  1995, ApJ, 442, 228

\bibitem[\protect\citeauthoryear{{Sod}}{{Sod}}{1978}]{sod78}
{Sod} G.~A.,  1978, J. Comput. Phys., 27, 1

\bibitem[\protect\citeauthoryear{{Stasyszyn}, {Dolag} \& {Beck}}{{Stasyszyn}
  et~al.}{2013}]{sdb12}
{Stasyszyn} F.~A.,  {Dolag} K.,    {Beck} A.~M.,  2013, MNRAS, 428, 13

\bibitem[\protect\citeauthoryear{{Stone}, {Gardiner}, {Teuben}, {Hawley} \&
  {Simon}}{{Stone} et~al.}{2008}]{athena}
{Stone} J.~M.,  {Gardiner} T.~A.,  {Teuben} P.,  {Hawley} J.~F.,    {Simon}
  J.~B.,  2008, ApJS, 178, 137

\bibitem[\protect\citeauthoryear{{T{\'o}th}}{{T{\'o}th}}{2000}]{toth2000}
{T{\'o}th} G.,  2000, J. Comput. Phys., 161, 605

\bibitem[\protect\citeauthoryear{{Tricco} \& {Price}}{{Tricco} \&
  {Price}}{2012}]{tricco&price12}
{Tricco} T.~S.,  {Price} D.~J.,  2012, J. Comput. Phys., 231, 7214

\bibitem[\protect\citeauthoryear{{Wadsley}, {Veeravalli} \&
  {Couchman}}{{Wadsley} et~al.}{2008}]{wadsley08}
{Wadsley} J.~W.,  {Veeravalli} G.,    {Couchman} H.~M.~P.,  2008, MNRAS, 387,
  427

\end{thebibliography}

\end{document}